\documentclass[aps,prl,twocolumn,groupedaddress]{revtex4-1}

\usepackage{graphicx,amsmath, amssymb, amsthm, amsfonts}
\usepackage{array, hyperref}
\usepackage{listings}
\usepackage{fancyhdr}
\usepackage{verbatim}

\graphicspath{{./Figures/}}

\begin{document}
\title{Split Fermi seas in one-dimensional Bose fluids}

\author{T. Fokkema}
\email{T.B.Fokkema@uva.nl}
\affiliation{Institute for Theoretical Physics, University of Amsterdam, Science Park 904\\
Postbus 94485, 1090 GL Amsterdam, The Netherlands}

\author{I. S. Eli\"ens}
\affiliation{Institute for Theoretical Physics, University of Amsterdam, Science Park 904\\
Postbus 94485, 1090 GL Amsterdam, The Netherlands}

\author{J.-S. Caux}
\affiliation{Institute for Theoretical Physics, University of Amsterdam, Science Park 904\\
Postbus 94485, 1090 GL Amsterdam, The Netherlands}

\date{\today}

\begin{abstract}
For the one-dimensional repulsive Bose gas (Lieb-Liniger model), we study a special class of highly-excited states obtained by giving a finite momentum to subgroups of particles. These states, which correspond to `splitting' the ground state Fermi sea-like quantum number configuration, are zero-entropy states which display interesting properties more normally associated to ground states. Using a numerically exact method based on integrability, we study these states' excitation spectrum, density correlations and momentum distribution functions. These correlations display power-law asymptotics, and are shown to be accurately described by an effective multicomponent Tomonaga-Luttinger liquid theory whose parameters are obtained from Bethe Ansatz. The non-universal correlation prefactors are moreover obtained from integrability, yielding a completely parameter-free fit of the correlator asymptotics.
\end{abstract}

\maketitle

\section{Introduction}
Many-body quantum physics in one dimension \cite{GiamarchiBOOK} is a well-known theater in which strong correlation effects systematically take the leading role. Besides providing many examples of quantum critical ground states, one-dimensional (1d) systems also interestingly suffer from the breakdown of single-particle pictures, making calculations difficult but providing interesting physics through the appearance of new forms of quasiparticles with unconventional (fractionalized) statistics, dispersions and correlations. 

Much of our understanding of 1d systems stems from the existence of robust nonperturbative methods developed over the last decades. First and foremost, the concept of the Tomonaga-Luttinger liquid \cite{1981_Haldane_PRL_47,*1981_Haldane_JPC_14} and the technique of bosonization \cite{GogolinBOOK,1998_von_Delft_AP_7} have provided the consistent framework for describing the universal low-energy physics of these systems. On the other hand, the existence of isolated examples of exactly-solvable 1d models \cite{KorepinBOOK} whose wavefunctions can be obtained from Bethe Ansatz \cite{1931_Bethe_ZP_71} has opened up the door to many nonperturbative computations of physical properties of representative systems.

One seldom-exploited characteristic of the Bethe Ansatz is that, in marked contrast to bosonization, it provides exact wavefunctions for {\it any} state in the Hilbert space, irrespective of its energy. Besides allowing to consider {\it e.g.} finite-temperature thermodynamics of exactly-solvable models, this fact also opens the door to the investigation of many more general issues going beyond conventional equilibrium physics.

Our aim here is to consider a relatively simple class of states of 1d repulsive bosons which display a number of interesting properties. These states, which can be intuitively pictured as eigenstates in which the fluid contains a number of `pockets' of atoms moving at distinct momenta, share many features with the ground state, including signs of quantum criticality. Since they are extremely highly excited states, they would ultimately be unstable to external perturbations; on the other hand, being eigenstates of a physically meaningful Hamiltonian, their lifetimes can in principle be extremely large if perturbations are weak. They therefore realize another instance of `metastable criticality' \cite{2013_Panfil_PRL_110} in interacting gases. 

On the experimental side, there has been remarkable progress in the realization and investigation of isolated quantum systems \cite{2008_Bloch_RMP_80}, in particular bosonic systems in 1d, these providing rich and interesting physics \cite{2011_Cazalilla_RMP_83}. Part of our motivation thus also came from the famous quantum Newton's cradle experiment \cite{2006_Kinoshita_NATURE_440} in which a Bragg pulse is used to produce an initial state with a doubly-peaked momentum distribution. In attempts to understand this experiment, proximity to integrability is often invoked as the physical reason for the absence of relaxation. Although not meant to model this initial state accurately, the states we consider do contain similar features in their momentum distribution function, and might form a useful starting point for a more refined attempt at phenomenology.

The paper is organized as follows. After introducing the Lieb-Liniger model, we precisely define the type of states that we will study. A detailed discussion of the excitations that govern the correlations in the system is followed by results for dynamical correlations obtained numerically from an integrability-based method. We provide results for the most easily measured observables, namely the density correlations (the dynamical structure factor) and the momentum distribution function. We then discuss how the system can be understood in terms of a multicomponent Luttinger liquid, for which all effective parameters are related to data computable from integrability. Finally, we compare the computed correlations with the asymptotic Luttinger liquid, parameter-free predictions, and end with conclusions and perspectives.

\section{Lieb-Liniger bosons}
The  Lieb-Liniger model \cite{1963_Lieb_PR_130_1, *1963_Lieb_PR_130_2} is a model for one-dimensional bosons with a $\delta$-function interaction. Setting $\hbar = 2m = 1$ the Hamiltonian of the model is
\begin{equation}
H = \int_0^L dx [\partial_x\Psi^\dag(x)\partial_x \Psi(x) +c\Psi^\dag(x)\Psi^\dag(x)\Psi(x)\Psi(x)]
\label{eq:LiebLinigerH}
\end{equation}
where $\Psi(x), \Psi^\dagger (x)$ are boson annihilation/creation operators obeying canonical equal-time commutation relations $[\Psi(x), \Psi^\dagger (x')] = \delta(x-x')$.
The interaction strength is parametrized by a single dimensionless parameter, $\gamma=c/ \rho_0$, where $\rho_0=N/L$ is the average density. We will only consider repulsive interactions $c>0$, and will for definiteness impose periodic boundary conditions. The $N$-particle wave functions are given by the Bethe ansatz as a linear combination of plane waves
\begin{equation}
\chi_N(x_1,\dots,x_N) = \prod_{j<k}^N \mathrm{sgn}(x_j-x_k) \sum_{P} \mathcal{A}_{P} e^{i \sum_j \lambda_{P_j} x_j},
\end{equation}
where $\mathcal{A}_P = (-1)^{[P]} e^{\frac{i}{2} \sum_{j>k} \mathrm{sgn}(x_j-x_k) \theta(\lambda_{P_j} - \lambda_{P_k})}$. The two-particle phase shifts are
$\theta(\lambda) = 2\arctan(\lambda/c)$ and $P$ denotes a permutation of $N$ indices.
With periodic boundary conditions, the eigenstates of the Lieb-Liniger Hamiltonian with $N$ particles are specified by giving $N$ quantum numbers which are integers for $N$ odd and half-odd-integers for $N$ even. Multiple occupation of quantum numbers is not allowed. In particular, the ground state is given by the quantum numbers $I \in \{-(N-1)/2, \ldots, (N-1)/2\}$, i.e. a Fermi sea configuration.
The quasi-momenta $\{\lambda_j\}_{j=1}^N$ relate to the quantum numbers $\{I_j\}_{j=1}^{N}$ through the Bethe equations as
\begin{equation}
 \lambda_j L= 2 \pi I_j - \sum_{l=1}^{N} \theta(\lambda_j - \lambda_l).
\end{equation}
The total momentum of an eigenstate is $P=\sum_{j=1}^N \lambda_j$ and is zero for the ground state. The energy of an eigenstate is  $E=\sum_{j=1}^N \lambda_j^2 $.

In the so-called Tonks-Girardeau limit  $c\to\infty$ \cite{1960_Girardeau_JMP_1, 2004_Paredes_NATURE_429}, the system simplifies considerably, reducing to a  gas of impenetrable bosons which, up to particle statistics, is equivalent to free fermions. 
We will use this limit later on as a separate check of our results.

\subsection{Zero-entropy critical states}
We will consider states in the Lieb-Liniger model corresponding to one or more intervals of  occupied quantum numbers (no holes) and all other quantum numbers unoccupied. The simplest case is the ground state, corresponding to a single interval as described above. In general, we can think of a number of disjoint intervals $[I_{1L},I_{1R}],\ldots,[I_{nL},I_{nR}]$. We will focus on the case of two intervals $[I_{1L},I_{1R}]$ and $[I_{2L},I_{2R}]$ in particular. As this type of state can be obtained by splitting the Fermi sea, we will generally refer to it as a \emph{Moses} state. 

Some immediate facts concerning Moses states are that their energy is thermodynamically large above that of the ground state, and their momentum distribution will be peaked around nonzero momenta (namely at momenta dictated by the size of and distance between the Fermi pockets), unlike the ground state case. Moreover, the fact that these states have zero entropy maximizes quantum effects in observables such as correlation functions. 

Such Moses states were previously considered in \cite{2011_Kormos_PRL_107} where their local two and three body correlations were calculated (similar correlations were computed in \cite{2011_Pozsgay_JSTAT_P01011,*2011_Kormos_PRA_83} using the method of \cite{2009_Kormos_PRL_103}). As argued in \cite{2011_Kormos_PRL_107}, the momentum kick caricatures the effects of the Bragg pulse performed in \cite{2006_Kinoshita_NATURE_440}. Of course the real experiment leads to a much more complicated initial state than a mere idealized Moses state such as considered here. One could however imagine defining a generalized Gibbs ensemble \cite{2008_Rigol_NATURE_452} with the power-law charges of the Lieb-Liniger model, as in \cite{2013_Eriksson_JPA_46}, specializing to a double-well generalized free energy leading to a description in terms of an ensemble of states in the vicinity of the Moses state. The Moses state would then be the zero-entropy limit of this GGE. Such a situation would differ from
an equilibrium situation where the ground state consists of a split Fermi sea, as occurs in certain phases of integrable spin ladder systems  \cite{Frahm:1999,*1997_Frahm_JPA_30,*2001_Zvyagin_EPB_19} where the dressed energy obtains a double well shape. The difference lies in the fact that in reality, two Hamiltonians have to be distinguished: the one setting the equilibrium/ensemble average, and the one driving the unitary quantum time evolution. In equilibrium these operators are identical. In a GGE (or similar) description out of equilibrium however, this is not the case. This distinction does not matter for static quantities (like the ones considered in \cite{2011_Kormos_PRL_107}), but does matter for dynamic ones as we study here, in which the Lieb-Liniger Hamiltonian (\ref{eq:LiebLinigerH}) drives the time evolution, but the Moses state emerges as the saddle-point state of some (here left unspecified) effective theory.

\begin{figure}
\includegraphics[width=7cm]{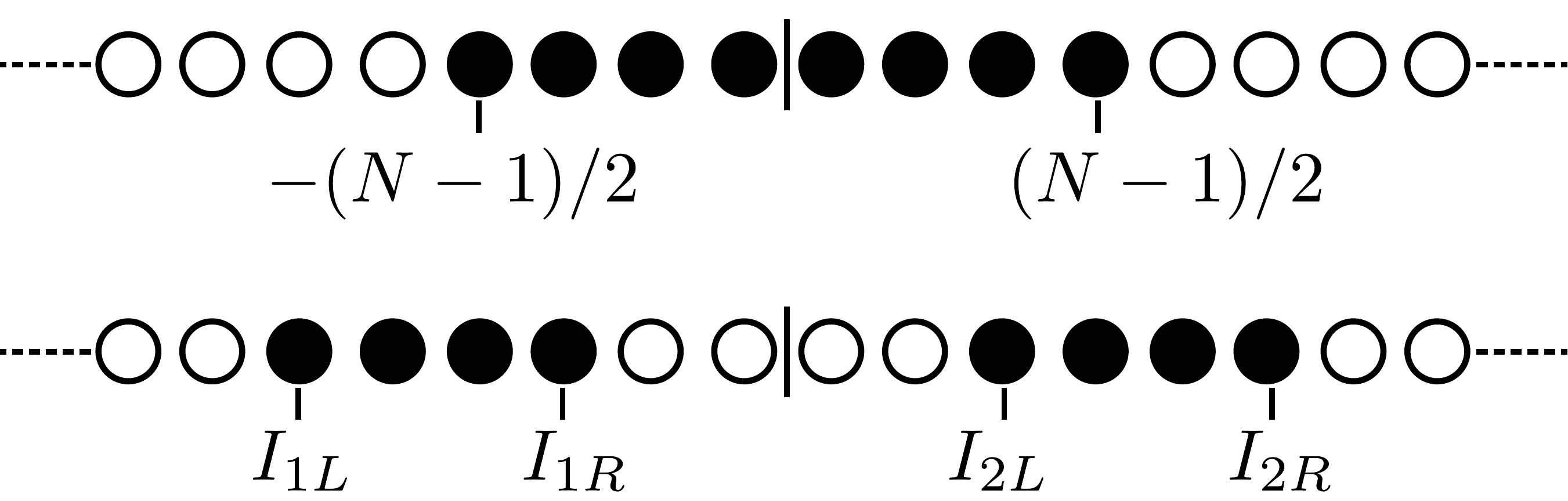}
\caption{\label{fig:moses_state}Illustration of the notations used to specify a Moses state. Top: The ground state has quantum numbers $-(N-1)/2,\dots,(N-1)/2$ occupied. Bottom: The Moses state specified by $\{I_{1L},I_{1R},I_{2L},I_{2R}\}$ has occupied quantum numbers in the intervals $[I_{1L},I_{1R}]$ and $[I_{2L},I_{2R}]$.}
\end{figure}

Let us establish some notations. We specify the particular Moses state under study by listing  the four extremal quantum numbers $\{I_{1L},I_{1R},I_{2L},I_{2R}\}$  or the associated `Fermi momenta' $\{k_{1L}, k_{1R}, k_{2L}, k_{2R}\}$ with $k_{ia} = \frac{2\pi}{L} I_{ia}$. We will use indices $i,j,k,\ldots = 1,2$ to denote the two `seas' and indices $a,b,c,\dots = L,R$ to denote the edge of a sea. The extremal quantum numbers $I_{ia}$ are mapped by the Bethe equations to the quasi-momenta $\lambda_{ia}$ which become equal to  $k_{ia}$  in the Tonks-Girardeau limit. It is useful to define $k_F=\sum_{ia} s_a k_{ia}= \pi \rho_0$ with $s_{R/L} = \pm 1$.  Fig.~\ref{fig:moses_state} illustrates the construction.

\section{Excitation spectrum}

\begin{figure}
\includegraphics[width=8.5cm]{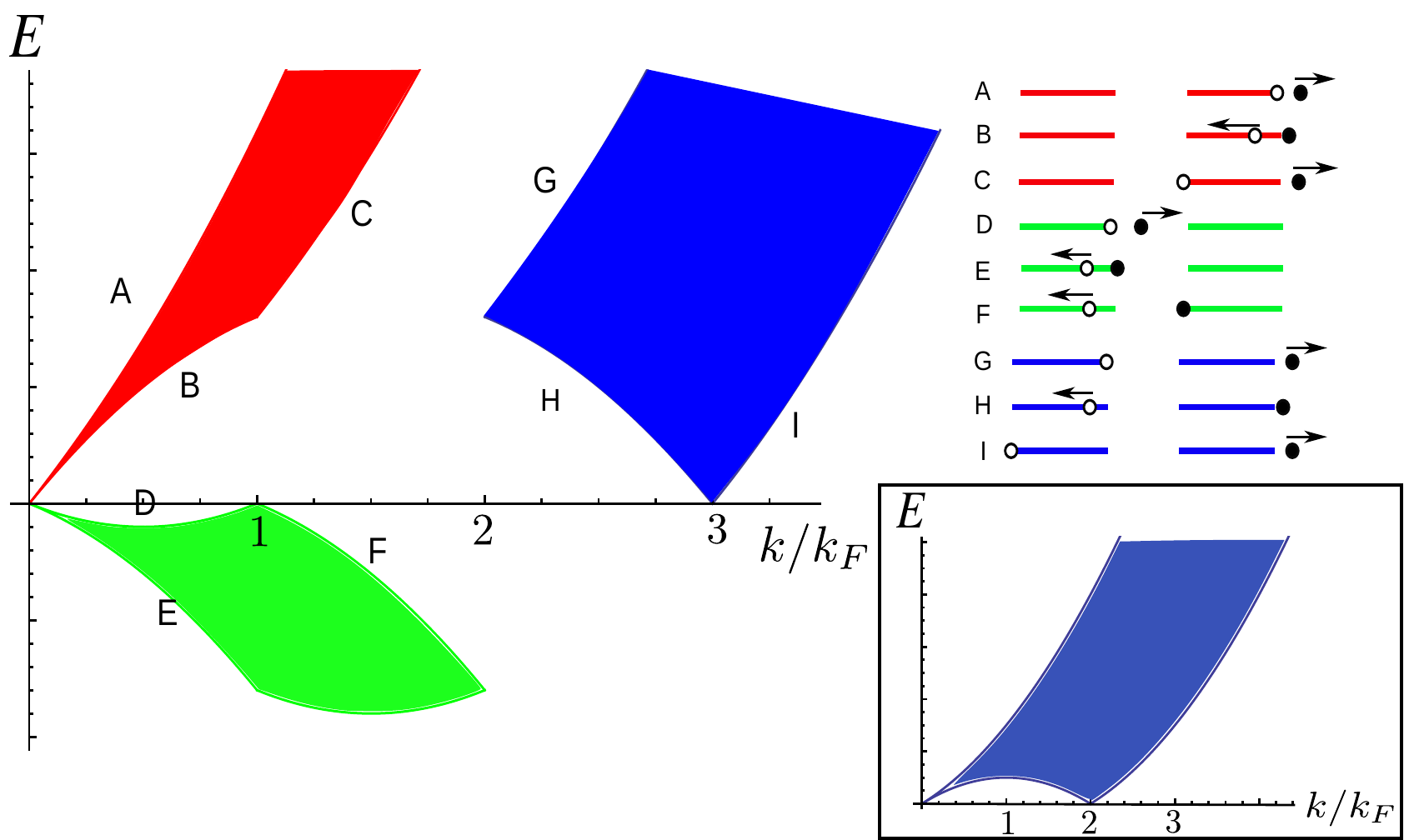}
\caption{\label{fig:spectrum_sym}(Color online.) The single particle-hole excitation spectrum for a Moses state at $c=\infty$ with Fermi momenta given by $\{-\frac{3}{2} \pi, -\frac{1}{2} \pi, \frac{1}{2} \pi, \frac{3}{2} \pi \}$ and a description of the excitations corresponding to the different lines.
 Starting with a particle excitation at $I_{2R}$ and a hole in the right Fermi sea we get line A, which corresponds to the standard Lieb-type-I excitation and line B which corresonds to the Lieb-type-II excitation. At the end of line B, when the hole is at $I_{2L}$, the hole cannot move any further to the left. Instead we can now have a particle in between the two Fermi seas and a hole at $I_{1R}$, this gives line D. Line H can be seen as a continuation of the Lieb-type-II excitation of line B. The other dispersion lines are easily understood in a similar fashion.
Inset: The single particle-hole excitation spectrum for the ground state at $c=\infty$. In both Moses and ground states, the single particle-hole continuum gives the dominant support for density correlations.}
\end{figure}

The splitting up of the Fermi sea quantum number configuration has great consequences on the structure of the excitation spectrum of the theory. For the case of the ground state, one can identify particle (Type I) and hole (Type II) branches \cite{1963_Lieb_PR_130_2} of soliton-like excitations \cite{2012_Sato_PRL_108}, leading to a characteristic single particle-hole continuum (see inset of Fig. \ref{fig:spectrum_sym}) clearly visible in correlation functions. The generalization of these modes to a symmetric Moses state is illustrated in Figure~\ref{fig:spectrum_sym}. The edges correspond to the new particle and hole dispersion lines generalizing the Lieb type-I and type-II modes. Due to the vacancies for quantum numbers in between the seas, part of the spectrum is shifted to  the negative energy domain, leaving a characteristic excluded area as compared to the ground state case. The types of excitations that correspond to the different parts of the spectrum are also indicated.  
Another difference compared to the ground state situation is that because of the negative energy branch the spectrum will become completely gapless if multiple particle-hole excitations are taken into account. We will discuss the spectrum and its linearization further when presenting the effective Tomonaga-Luttinger description of states in the vicinity of the Moses states.

\begin{figure}
\includegraphics[width=4.25cm]{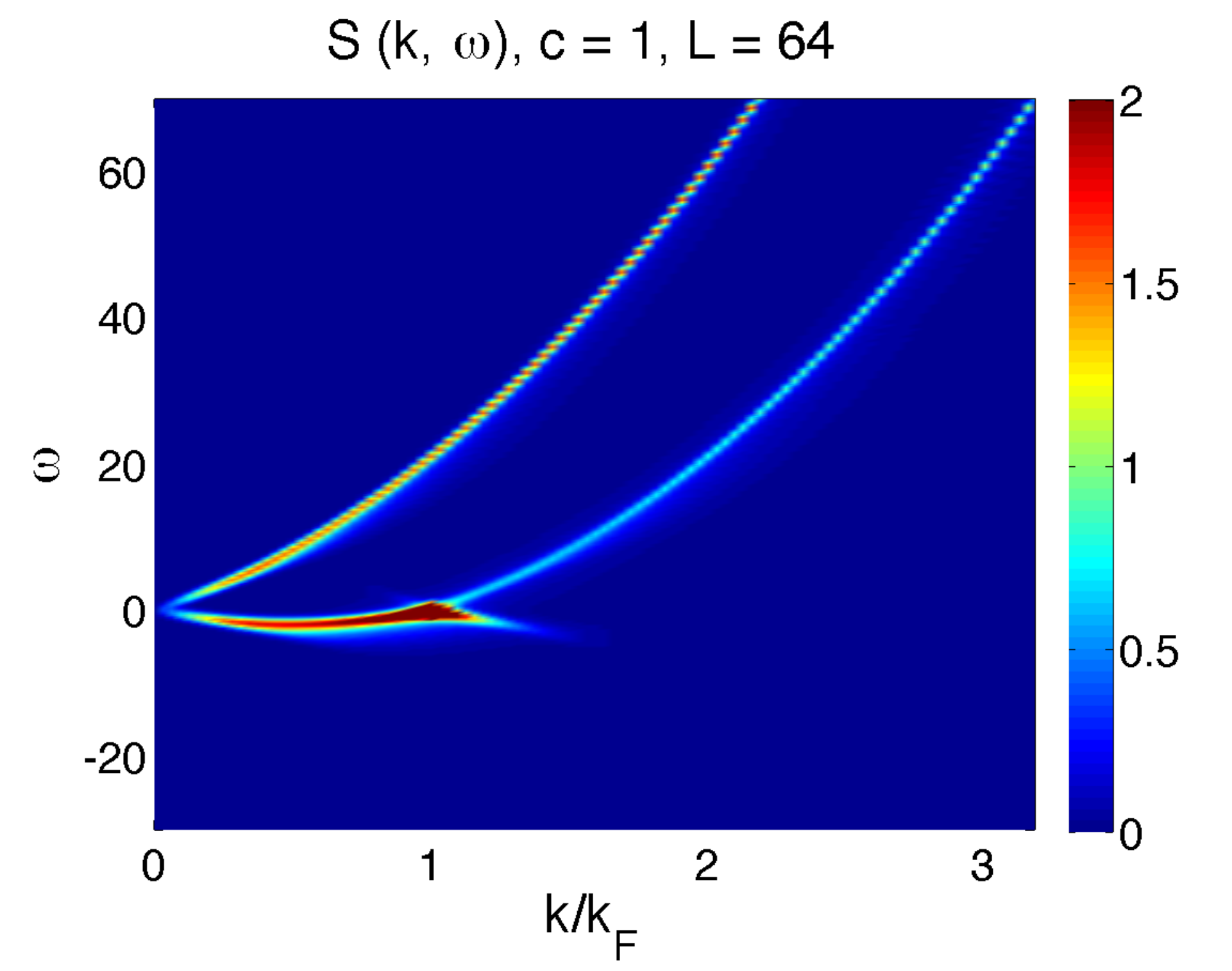}
\includegraphics[width=4.25cm]{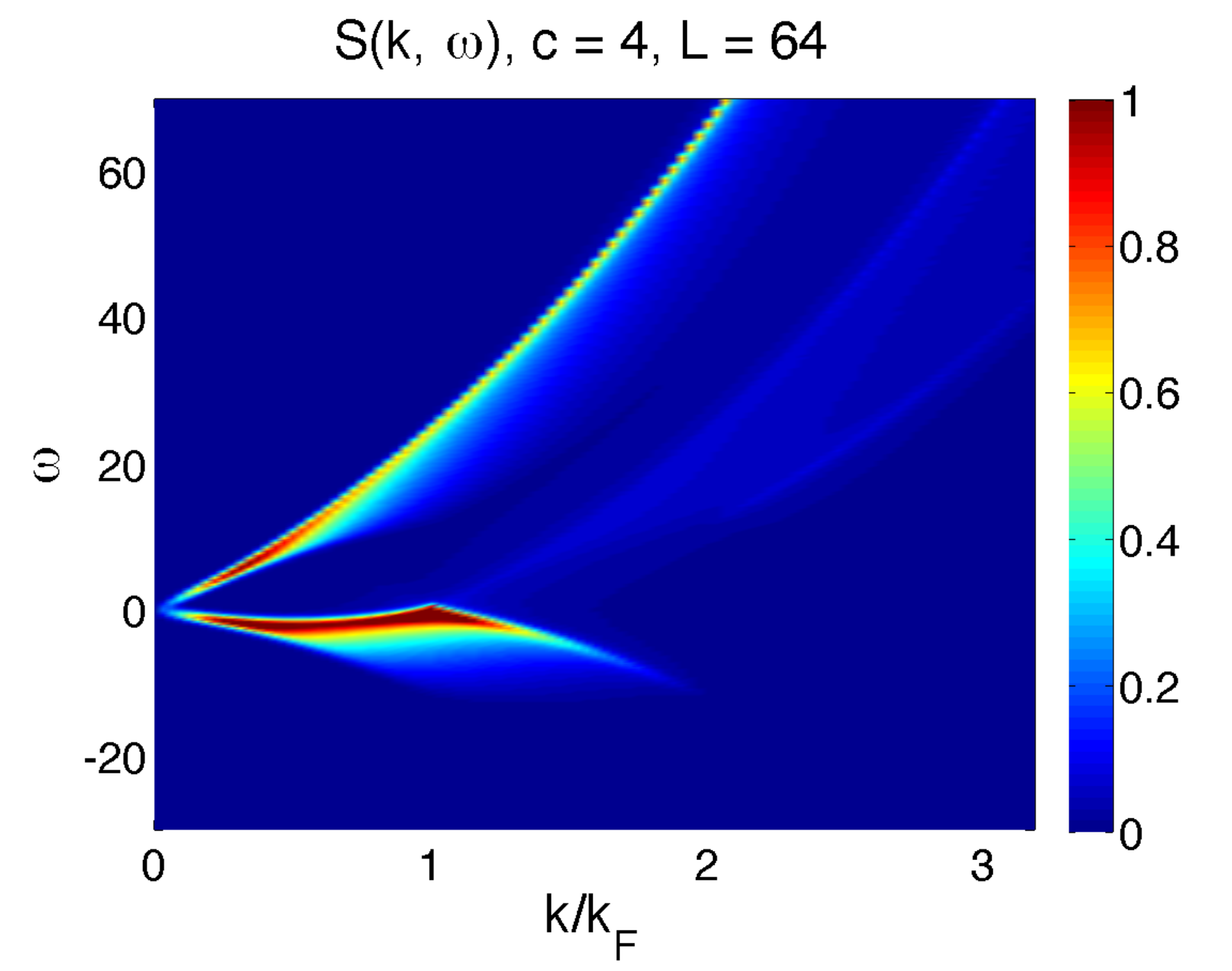}
\includegraphics[width=4.25cm]{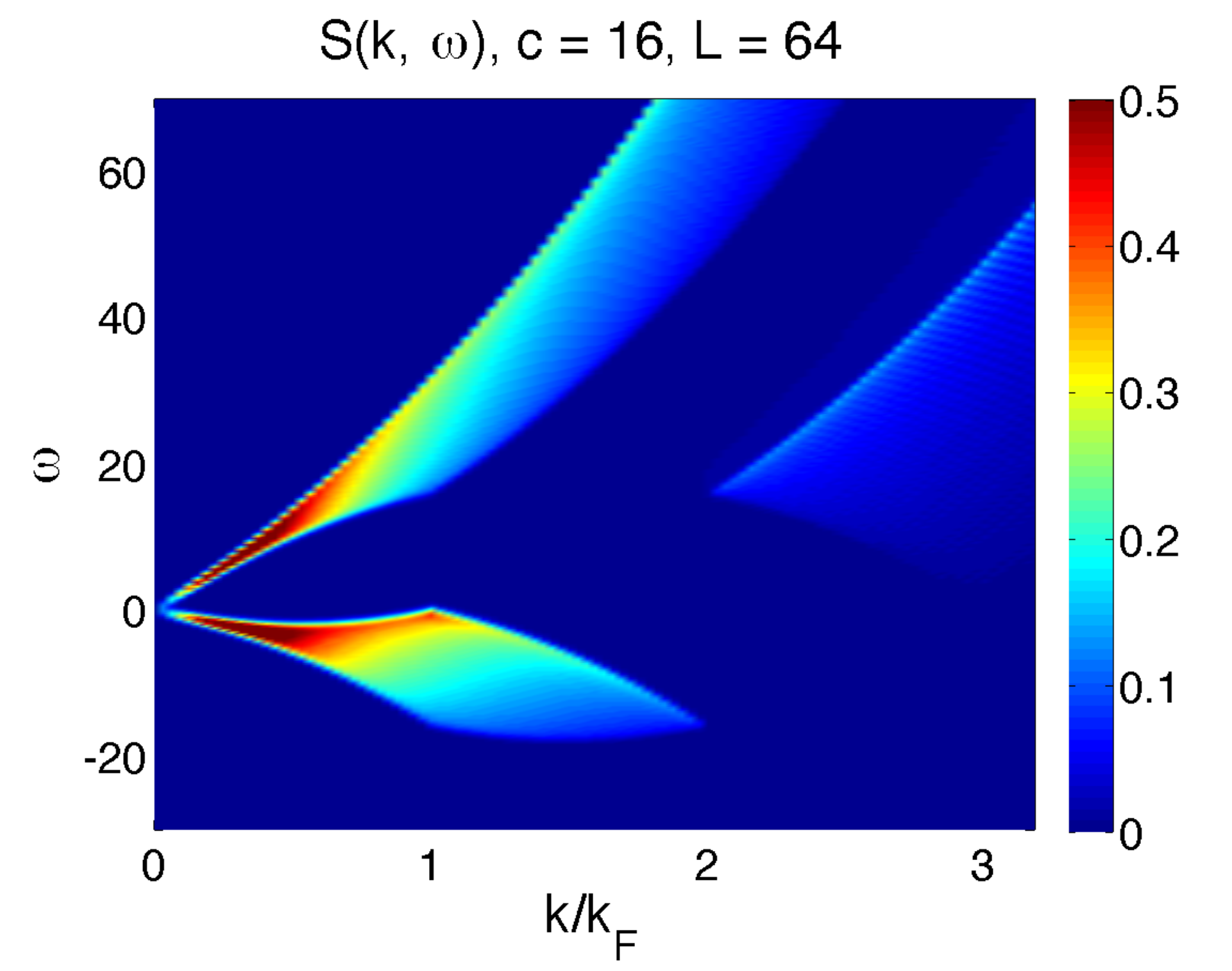}
\includegraphics[width=4.25cm]{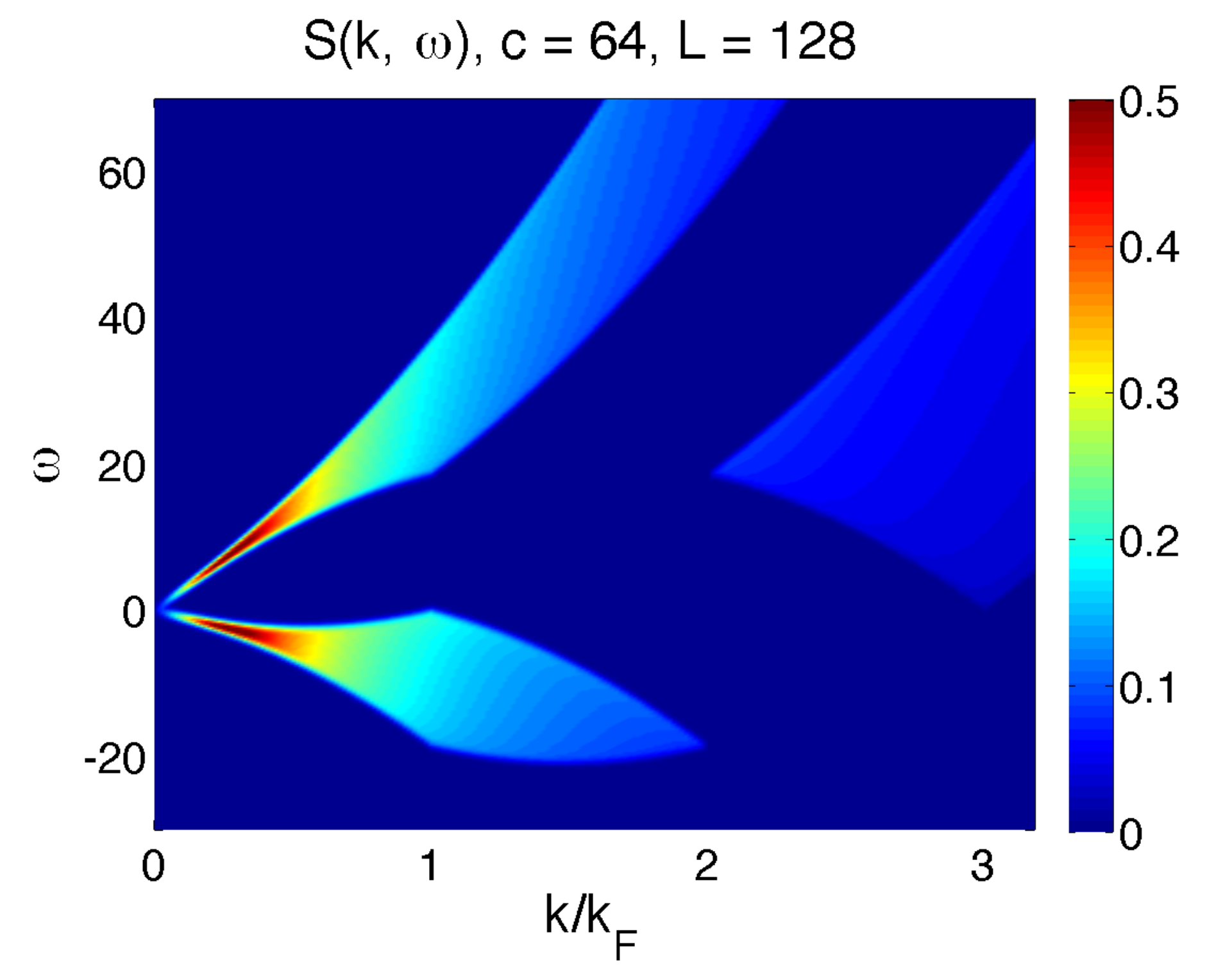}
\caption{\label{fig:dsf}(Color online.) The dynamical structure factor $S(k,\omega)$ for the Moses state for different values of the interaction strength. From top left to bottom right: $c=1, c=4,c=16,c=64$. The calculation was performed for 64 particles,  32 in each Fermi sea, and 32 holes separating the two Fermi seas, except for $c=64$ where 128 particles were used.}
\end{figure} 

\section{Correlation functions from integrability}
\subsection{Dynamical structure factor}
The dynamical structure factor (DSF) is defined as
\begin{align}
S(k,\omega) &= \int_{-\infty}^\infty dt \int_0^L dx e^{i\omega t-ikx} \langle \rho(x,t) \rho(0,0)\rangle\nonumber\\
&= \frac{2 \pi}{L} \sum_\alpha |\langle M | \rho_k |\alpha\rangle |^2 \delta(\omega - E_\alpha + E_0),
\end{align}
where $| M \rangle$ symbolizes the Moses state, $\alpha$ labels a complete set $\{|\alpha \rangle\}$ of eigenstates with energies $E_\alpha$, and $\rho(x) = \Psi^\dag(x)\Psi(x)$ is the density operator. This correlator is an efficient probe of the structure of particle-hole excitations. It relates directly to the linear response of the system with respect to perturbations coupling to the density. We have used the ABACUS routine \cite{2009_Caux_JMP_50} to  evaluate the DSF numerically (see Fig.~\ref{fig:dsf}), generalizing the ground state DSF \cite{2006_Caux_PRA_74}. The DSF of Moses states displays a number of features paralleling those of the ground state. First of all, the vast majority of the correlation weight is located within the single particle-hole continuum. At the edges of this continuum, the DSF displays threshold singularities with interaction- and momentum-dependent exponents. Within the continuum, the distribution of correlation weight is strongly interaction-dependent. 

\begin{figure}
\includegraphics[width=8.5cm]{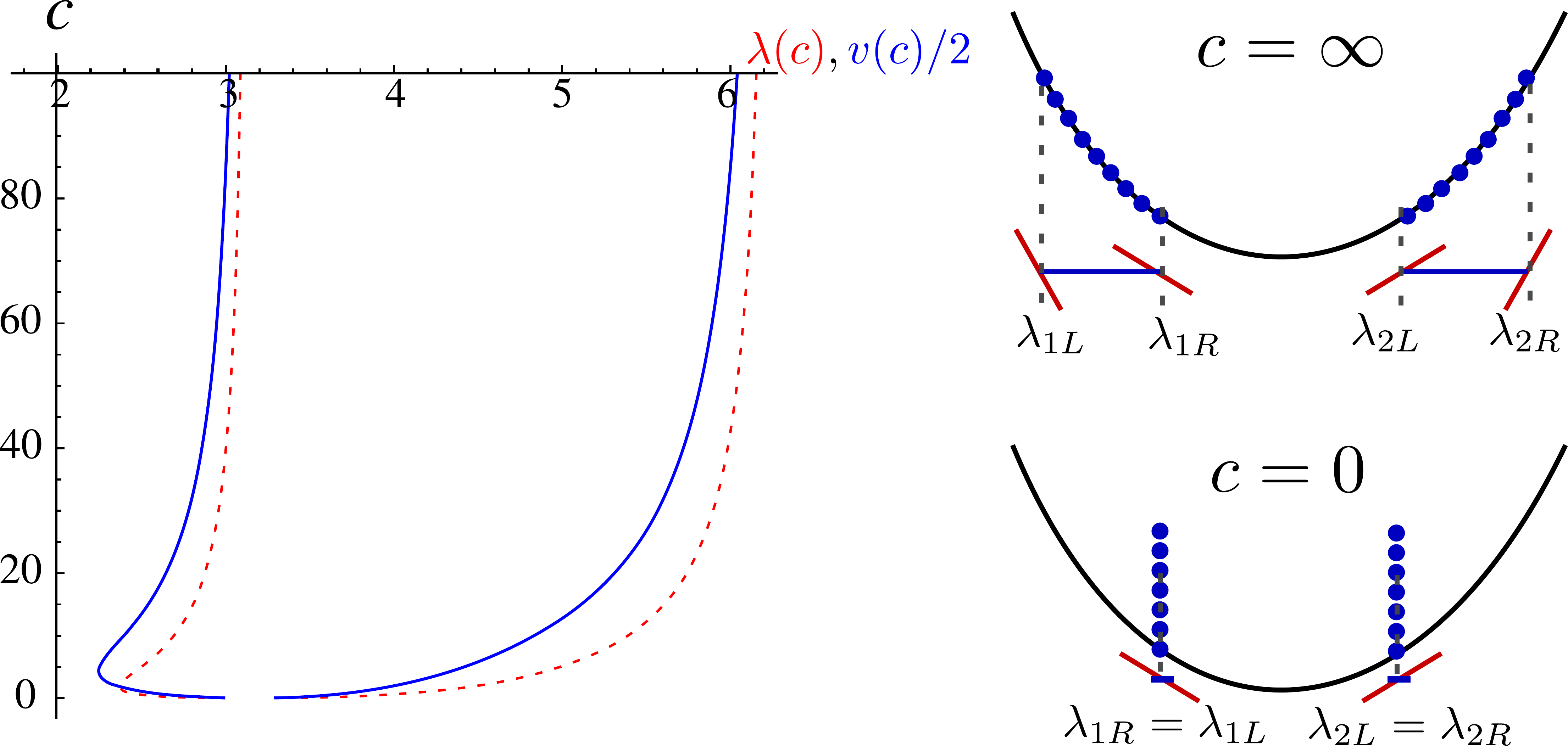}
\caption{\label{fig:fermispeed_rapidities}(Color online.) Left: Extremal rapidities $\lambda_{2L},\lambda_{2R}$  of the right Fermi sea (dashed red line) and velocities $\tilde{v}_{2L}/2,\tilde{v}_{2R}/2$ (solid blue line) as a function of $c$ (in units where $2m=1$) calculated for a state with Fermi momenta $k_{ia}=\{-2\pi, -\pi, \pi, 2 \pi\}$. Right: Illustration of the Moses state in the limits $c=\infty, 0$. In the $c=\infty$ limit, the extremal rapidities correspond to $\lambda_{ia} = k_{ia} = {2\pi I_{ia}}/{L}$. For $c=0$ the rapidities collapse onto the inner value $\lambda_{2L},\lambda_{2R}\to k_{2L}$. In the limits $c=0$ and $c=\infty$ the velocities are consistent with quadratic dispersion, i.e. $\tilde{v}_{ia} = 2 \lambda_{ia}$.}
\end{figure}

For small interactions the system becomes more and more like two coupled BECs as can also be seen from the solution of the extremal rapidities that collapse onto each other when $c\to0$ (see Fig.~\ref{fig:fermispeed_rapidities}). The DSF is then extremely sharply peaked at low energy and at a momentum corresponding to the distance between the internal edges of the two seas. On the other hand, for very large interactions, the DSF becomes essentially energy-independent, and its support espouses the single particle-hole continuum of Fig. \ref{fig:spectrum_sym}. 

\begin{figure}
\includegraphics[width=8cm]{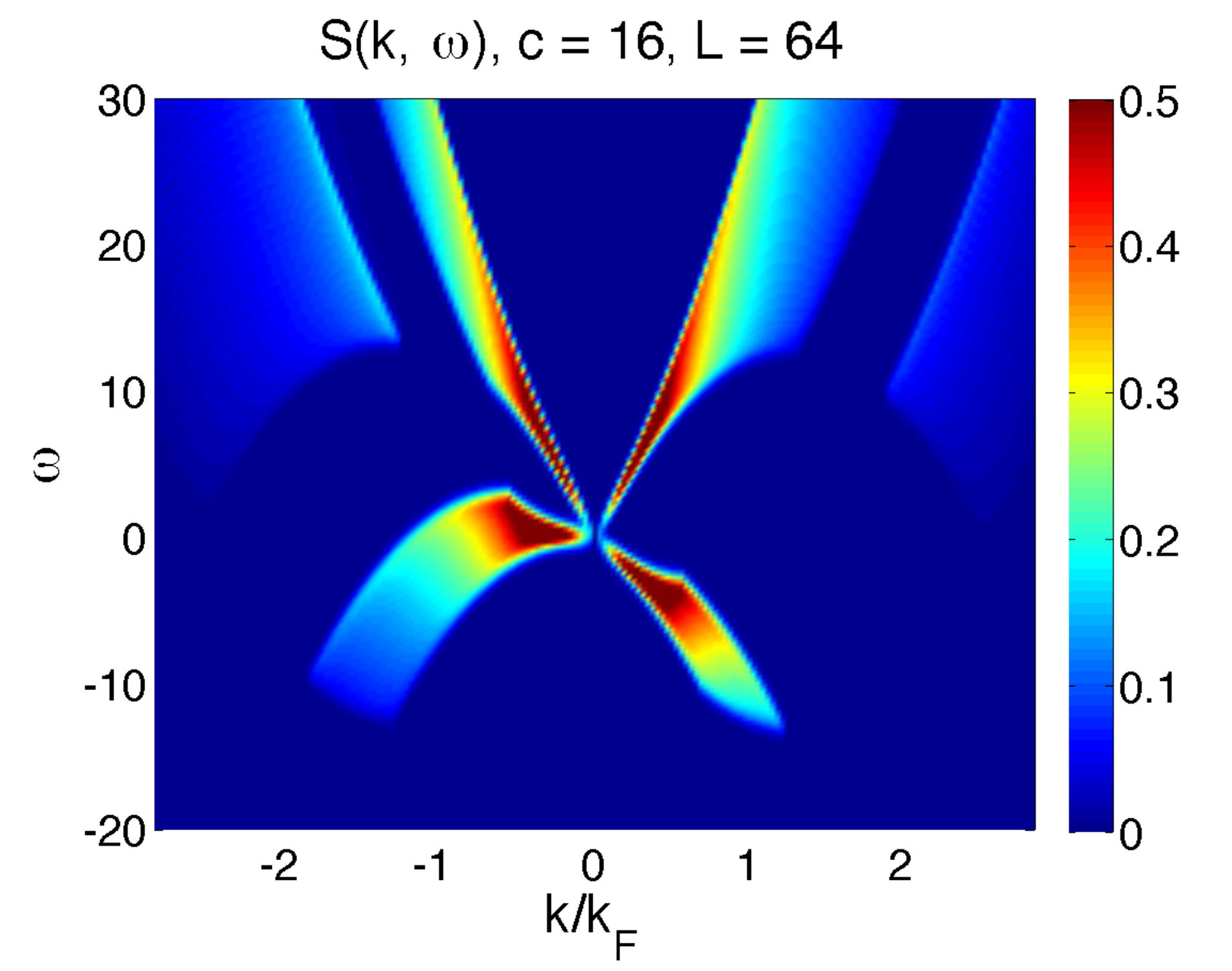}
\caption{\label{fig:dsf_asym}(Color online.) The dynamical structure factor $S(k,\omega)$ for an asymmetric Moses state for $c=16$. The calculation was performed for 64 particles with quantum numbers $\{I_{1L}, I_{1R}, I_{2L}, I_{2R}\}=\{-41.5, -20.5, -1.5, 39.5\}$.}
\end{figure}

In Fig.~\ref{fig:dsf_asym} the DSF for an asymmetric configuration is shown. The effects of `unbalancing' the Fermi pockets is quite easily visualized by following the changes in the dispersion lines. All features of the DSF mentioned above survive imposing such an asymmetry with minimal change.

\begin{figure}
\includegraphics[width=7cm]{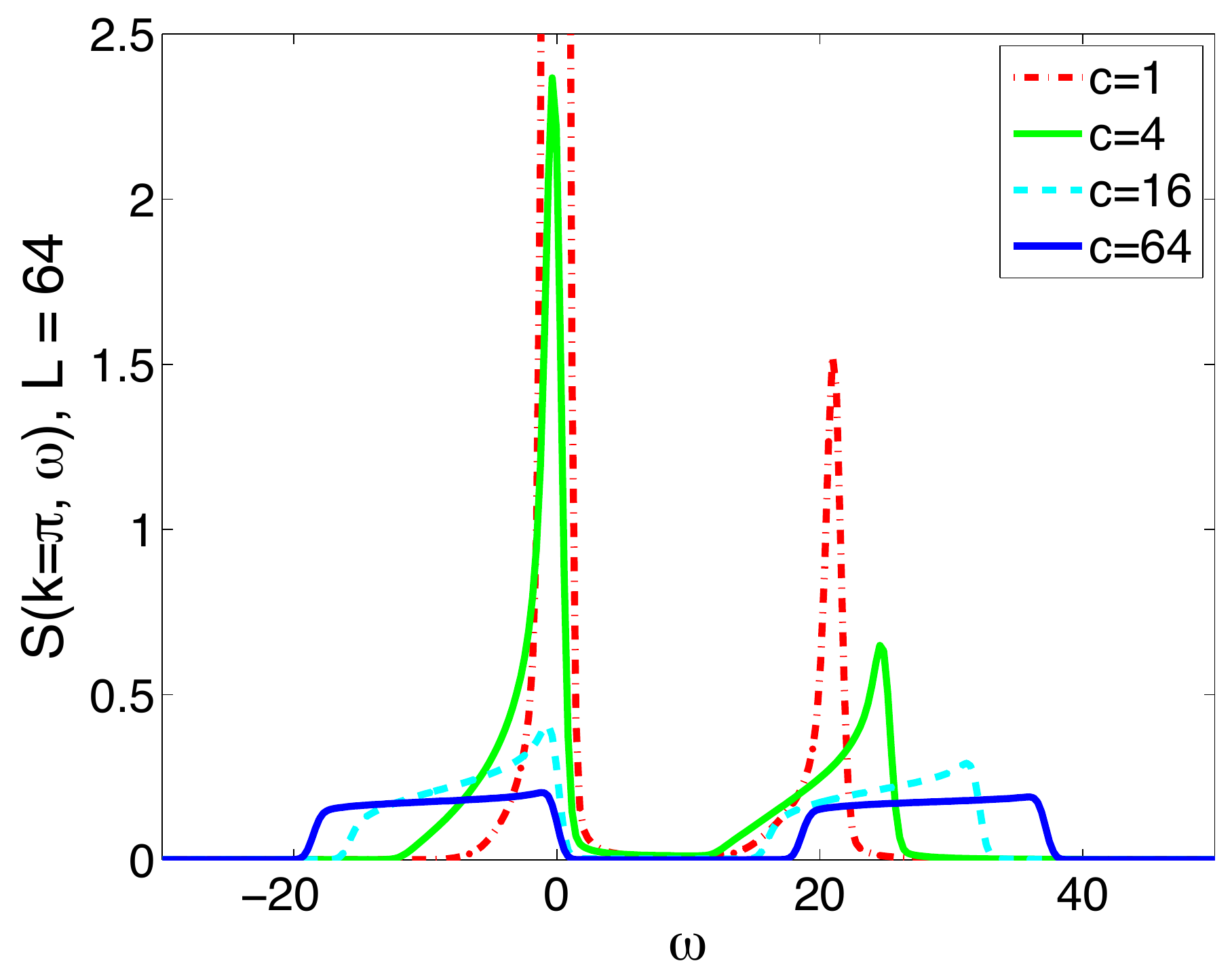}
\includegraphics[width=7cm]{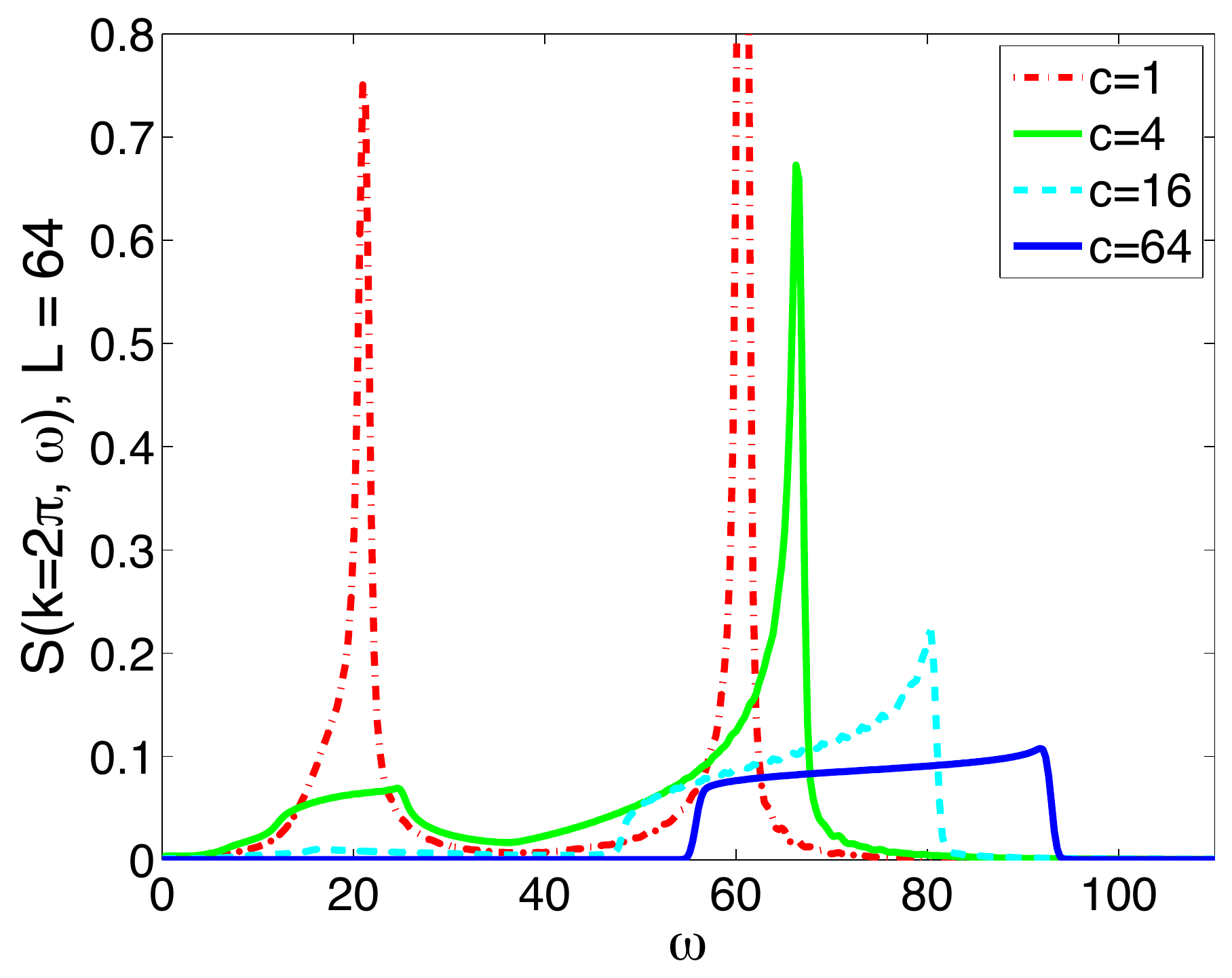}
\caption{\label{fig:dsf_cuts}(Color online.) Fixed momentum cuts of the dynamical structure factor $S(k,\omega)$ at $k=\pi$ (top) and $k=2\pi$ (bottom) for $c=1,4,16,64$. These graphs are obtained from the same data as in Fig. \ref{fig:dsf}, showing the threshold singularities at the edges of the single particle-hole continuum.}
\end{figure}

In Fig.~\ref{fig:dsf_cuts} momentum cuts of the DSF at fixed momenta $\pi$ and $2\pi$ are shown. The threshold singularities are clearly seen, as well as the flattening out of the correlation for increasing interaction strength. 

The quality of the computations is evaluated with sumrules. For the DSF, the f-sumrule $\int_{-\infty}^{\infty} \omega S(k, \omega) \frac{d\omega}{2\pi} = \frac{N}{L} k^2$ was used. Saturation levels at two representative momenta for all data sets presented in Fig. \ref{fig:dsf} are given in Table \ref{tab:Saturation}. Lower momenta are saturated better than the percentages given.

\begin{table}
\caption{\label{tab:Saturation}Levels of saturation of the f-sum rule for the DSF computations presented in Fig. \ref{fig:dsf}.}
\begin{ruledtabular}
\begin{tabular}{c | c  c  c  c}
& $c=1$ & $c=4$ & $c=16$ & $c=64$ \\
\hline
$k=\pi$ & 99.3\% & 99.2\% & 99.6\% & 99.7\% \\
$k=2\pi$ & 98.1\% & 97.0\% & 98.6\% & 98.9\% 
\end{tabular}
\end{ruledtabular}
\end{table}

\subsection{Static functions}
\begin{figure}
\includegraphics[width=8.5cm]{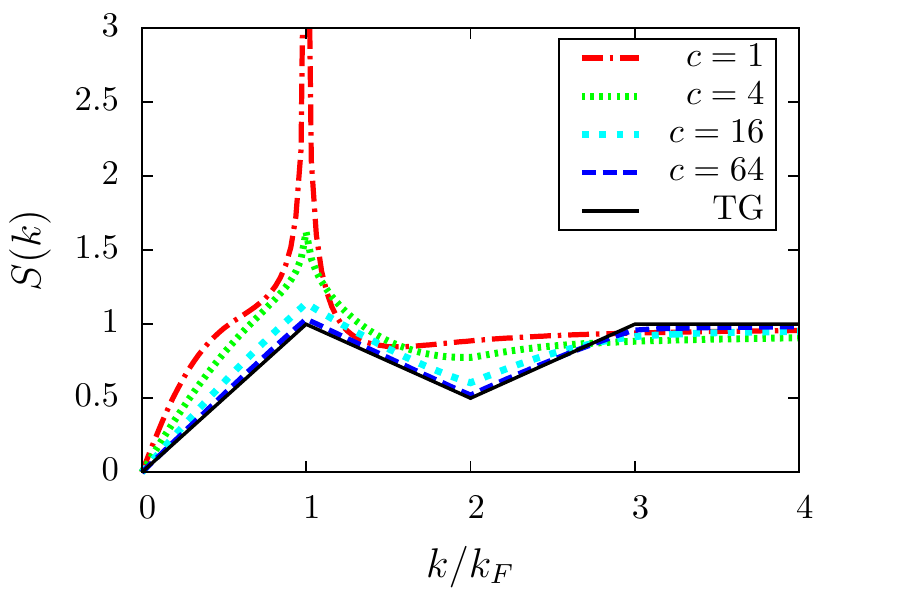}
\caption{\label{fig:sk2}(Color online.) The static structure factor for $c=1$, $c=4$, $c=16$ and $c=64$ with the exact Tonks-Girardeau result as benchmark. The data presented here is the frequency-integrated data of Fig. \ref{fig:dsf}.}
\end{figure}

\begin{figure}
\includegraphics[width=4.25cm]{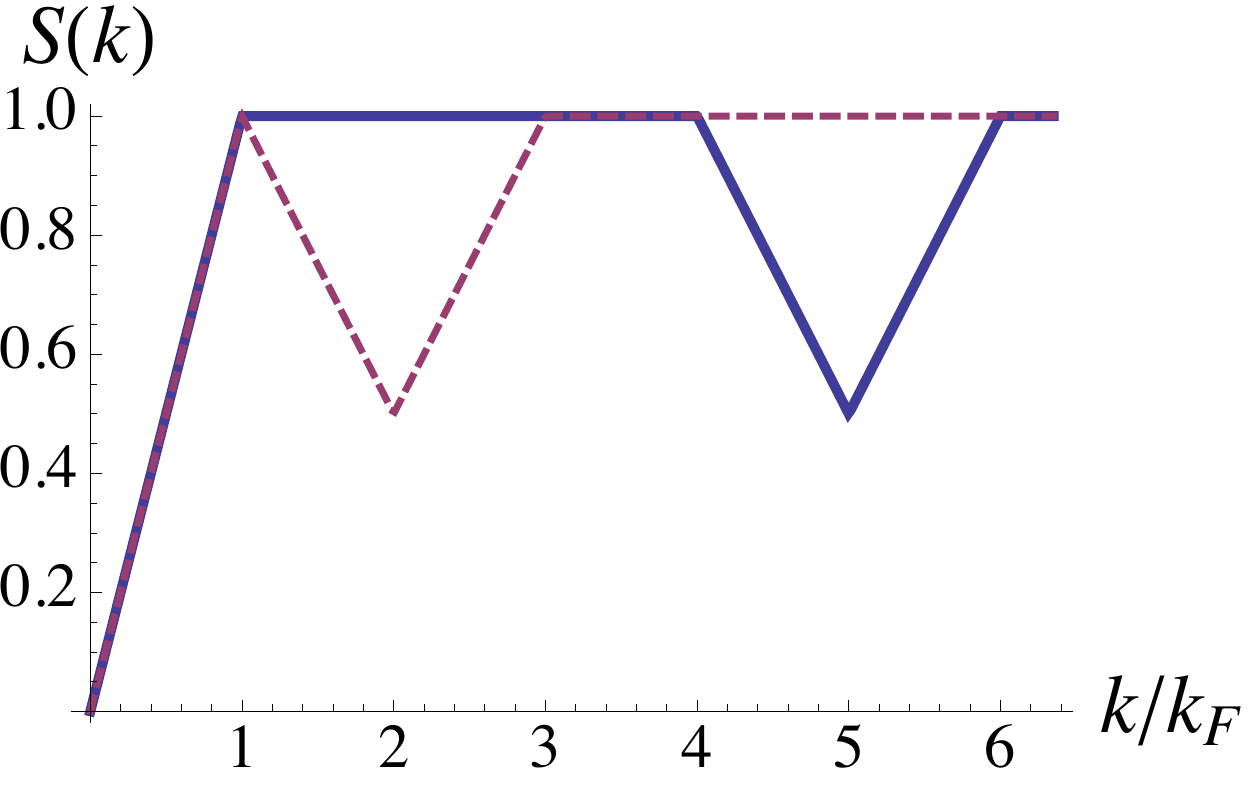}
\includegraphics[width=4.25cm]{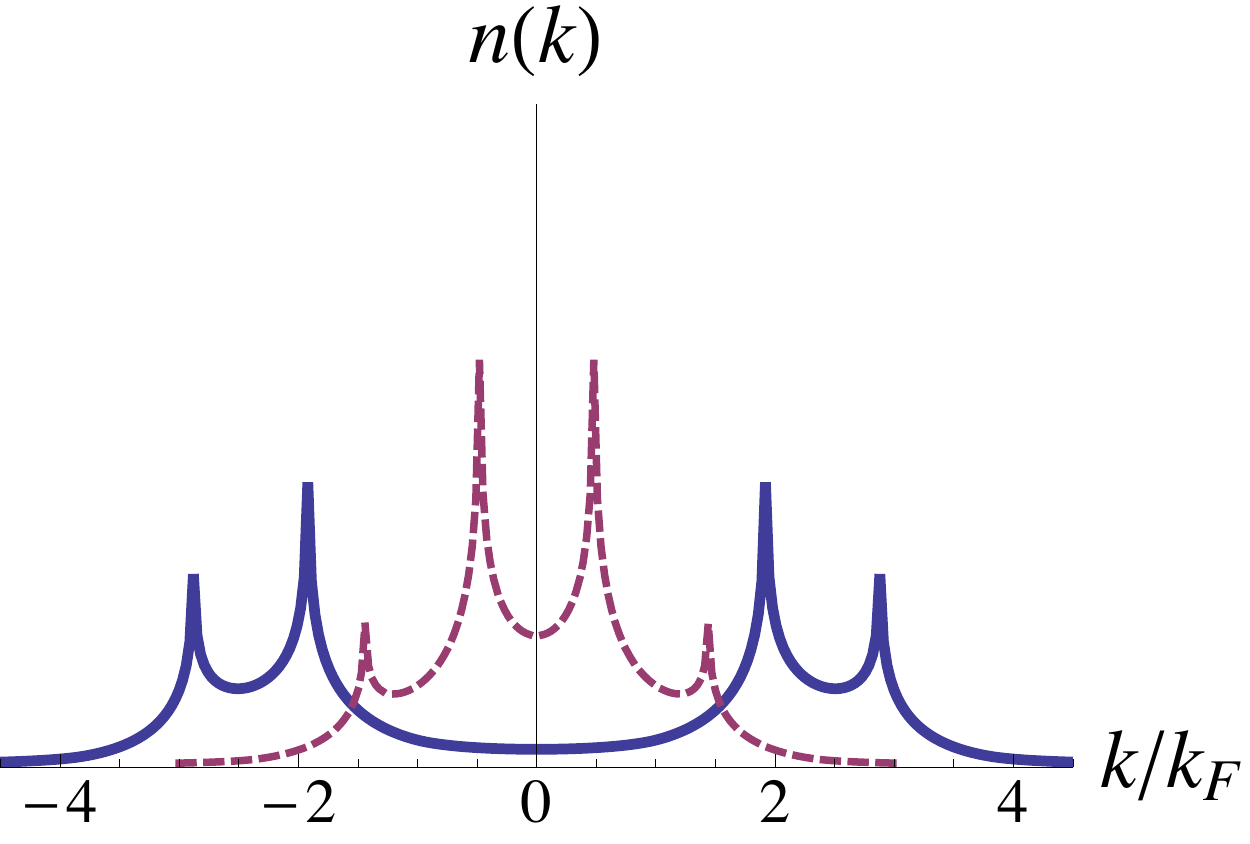}
\includegraphics[width=4.25cm]{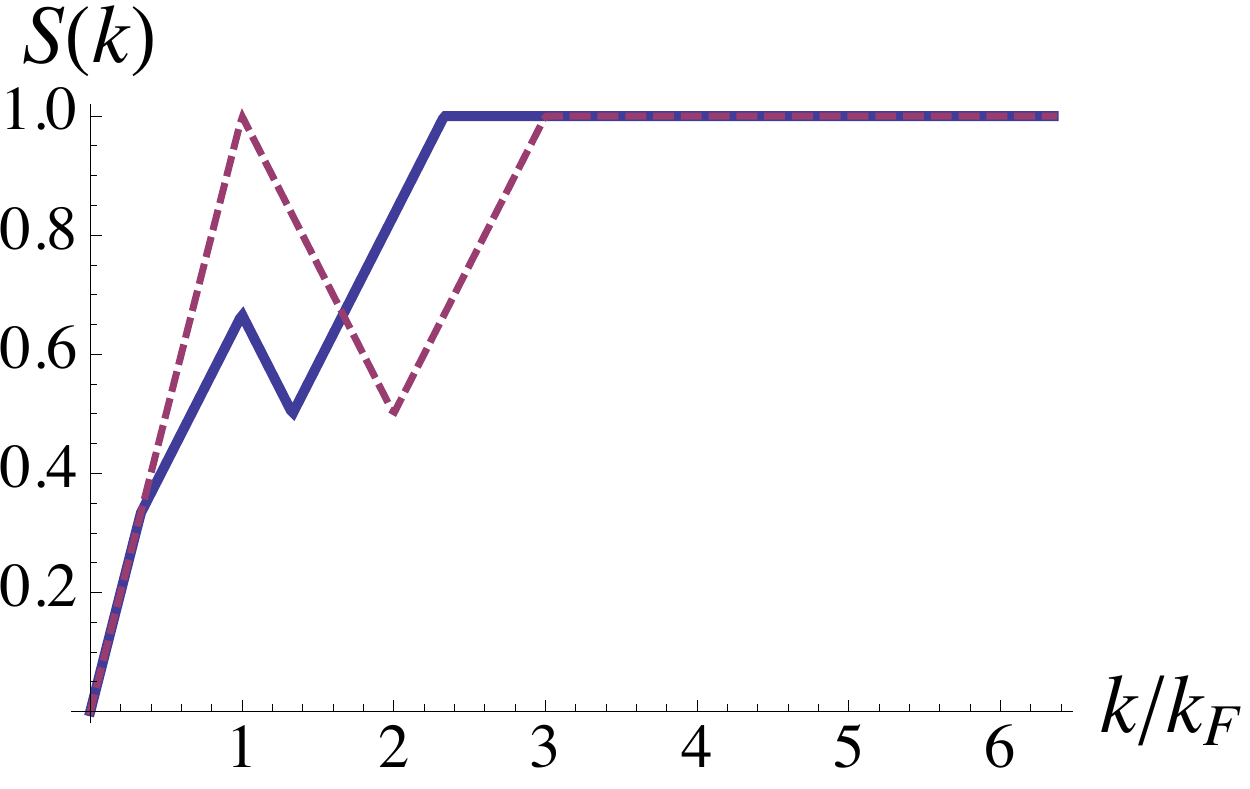}
\includegraphics[width=4.25cm]{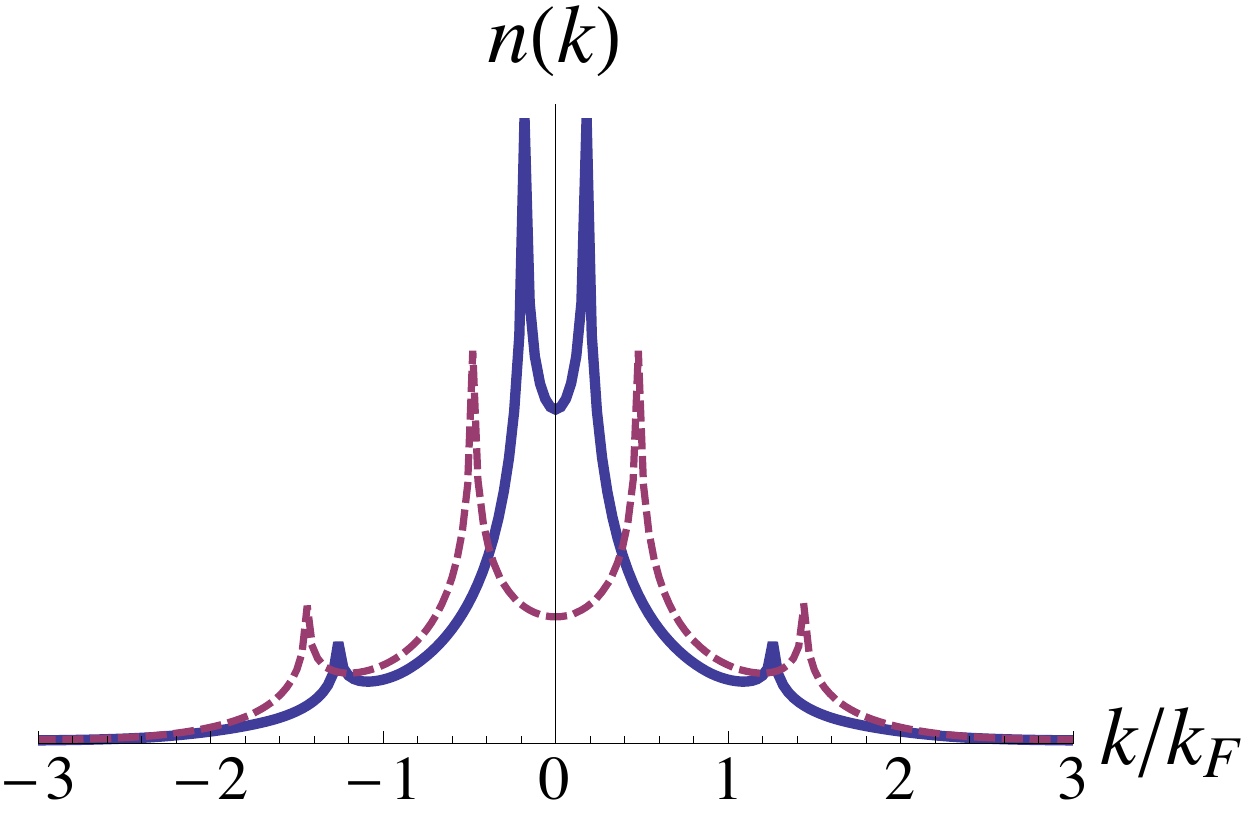}
\includegraphics[width=4.25cm]{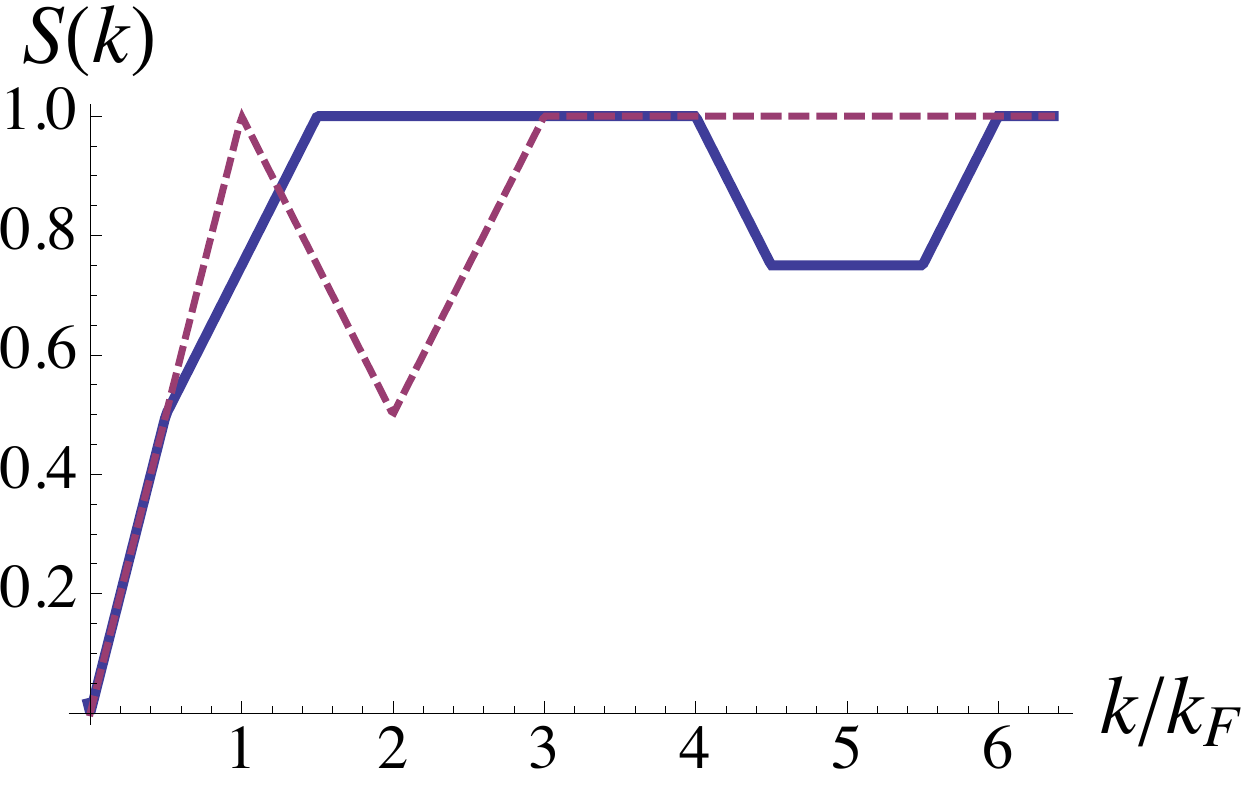}
\includegraphics[width=4.25cm]{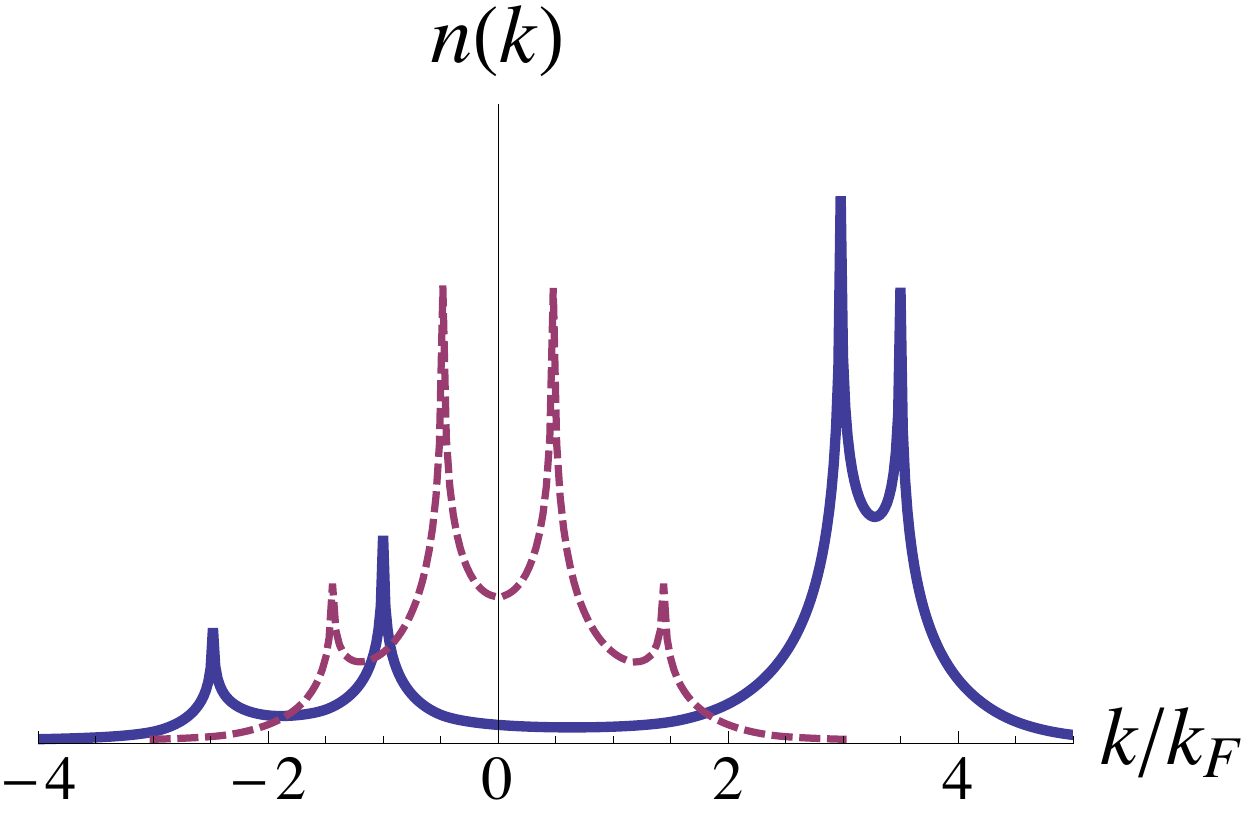}
\caption{\label{fig:sknk_compare}(Color online.) Illustration of the effect of changing the Moses state quantum number configuration on static correlation functions, using the solvable $c=\infty$ limit. The dashed lines represent a standard situation with Fermi momenta $\{-3/2 \pi, -1/2 \pi, 1/2 \pi, 3/2 \pi\}$, the solid lines correspond to a modified configuration. From top to bottom: increasing  separation between seas, decreasing separation between seas,  asymmetric seas with Fermi momenta $\{-2\pi, -3/2\pi, 5/2\pi, 4\pi\}$.  Left: The static structure factor. Right: Momentum distribution function.}
\end{figure}

\paragraph{Static structure factor.}
The dynamical functions give direct access to static correlations functions.
From the DSF, we obtain the static structure factor (SSF)
\begin{equation}
S(k) = \int \frac{d\omega}{2 \pi} S(k, \omega)
\end{equation}
which is plotted in Fig. \ref{fig:sk2} for different values of the interaction strength in a symmetric Moses state. Fig. \ref{fig:sknk_compare} shows the SSF easily obtained from single particle-hole excitations in the Tonks-Girardeau limit for different configurations, illustrating the effects of varying the momentum distance and configuration of the seas.

\paragraph{Momentum distribution function.}
The momentum distribution function (MDF) is defined as the Fourier transform of the static one-body function:
\begin{equation}
\label{eqn:momentumdistr}
n(k) = \int_0^L dx e^{ikx} \langle \Psi^\dagger (x) \Psi(0) \rangle.
\end{equation}
Let us begin by discussing the impenetrable limit, which illustrates some generic features and can be treated analytically following the work of Lenard \cite{1964_Lenard_JMP_5}. This calculation can also be done for states different from the ground state and we used the result in the form given in \cite{zvonarev_thesis}.
In Fig.~\ref{fig:sknk_compare}, the momentum distribution for different configurations of the Fermi seas is given in the Tonks-Girardeau limit. In this limit we see four peaks in the momentum distribution. If the Fermi momenta are given by $\{k_{1L}, k_{1R}, k_{2L}, k_{2R}\}$ then the peaks are located at $\{k_{1L}+k_F, k_{1R} - k_F, k_{2L} + k_F, k_{2R} - k_F\}$, where $k_F=\sum_{ia} s_a k_{ia}=\pi \rho_0$. The outer Fermi edges thus give rise to the inner peaks in the momentum distribution, and inner Fermi edges to the outer peaks. Moving the two seas closer to each other, we see that the outer peak becomes smaller and the inner peak becomes larger and closer to zero. If the Fermi seas are far away from each other the peaks become equally large. The limit of the two Fermi seas far away from each other is thus completely different from two non-interacting Fermi seas. In the latter case one would just have two peaks and not four. 

The MDF for generic values of the interaction parameter $c$ is computed using ABACUS again using an adaptation of the ground state algorithm used in \cite{2007_Caux_JSTAT_P01008}. For finite values of $c$, the outer peak of the momentum distribution becomes smaller and the inner peak larger, since reducing the repulsive interactions makes the system more like a decoupled BEC, whose MDF would have two isolated delta peaks. Already for $c \lesssim 10$, the momentum distribution shows only two distinguishable momentum peaks instead of four. One can thus view the internal peaks as `BEC'-driven, and the outer ones as `interaction'-driven. It is an interesting challenge for experiments  to try to create a state sufficiently similar to a Moses state in a tight toroidal trap \cite{2007_Ryu_PRL_99} such that the interaction-driven peaks in the MDF are visible, thereby demonstrating that the system is in a highly correlated quantum state,  similar to the phase correlations in  spatially split one-dimensional Bose gases \cite{2012_Gring_SCIENCE_14}.

\begin{figure}
\includegraphics[width=8.5cm]{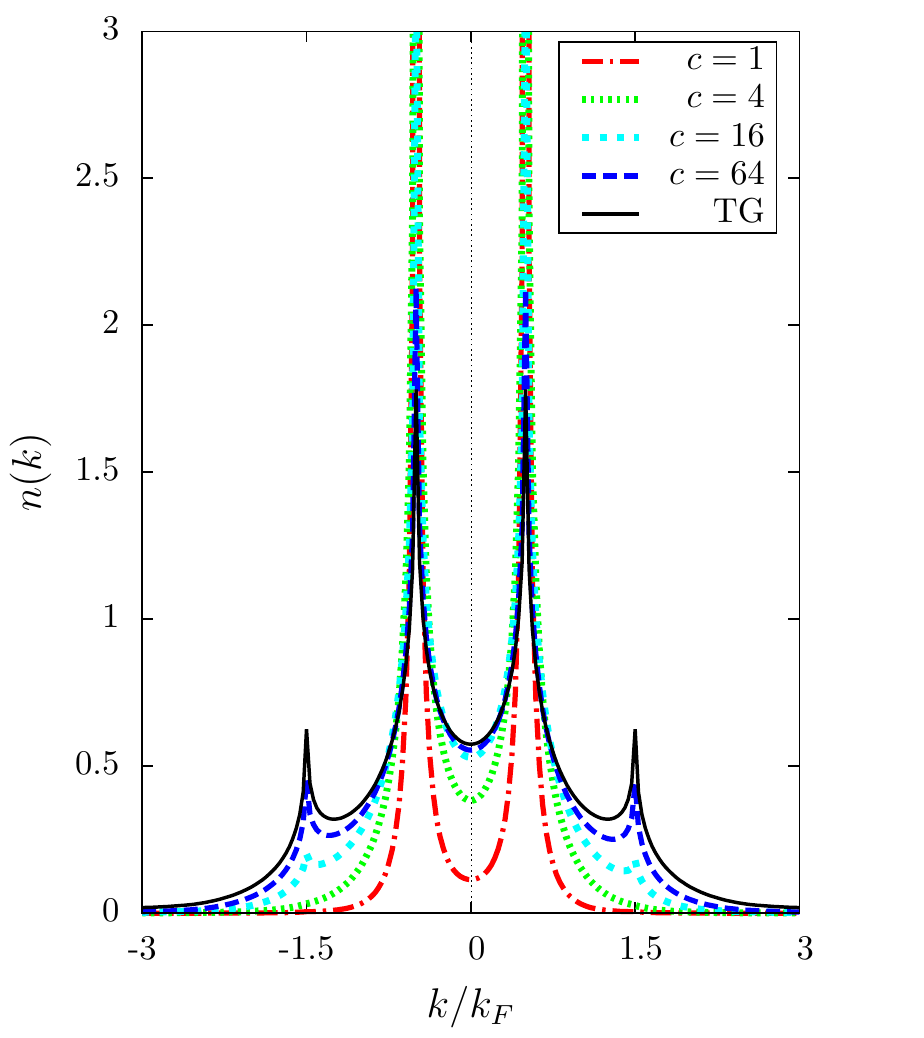}
\caption{(Color online.) The momentum distribution for different values of the interaction ($c=1,, c=4,\, c=16,\, c=64$) and the exact Tonks-Girardeau result. Calculations were performed on a Moses state with a symmetric configuration of filled quantum numbers: two Fermi seas of 32 particles each, separated by 32 holes.}
\label{fig:n_k} 
\end{figure}

\section{Multicomponent Tomonaga-Luttinger model description} 
One-dimensional quantum liquids in low-temperature equilibrium generally fall into the universality class of the Luttinger liquid \cite{1981_Haldane_PRL_47,1981_Haldane_JPC_14,GiamarchiBOOK}. The low-energy physics is dominated by excitations in the vicinity of the Fermi points $\pm k_F$. By linearizing the dispersion relation, an effective description in terms of free bosons can be obtained, parametrized by the sound  velocity $v$ and a single parameter $K$ encoding the interactions.

Although we study a far-from-ground state, its physical properties are governed by states that are, in  a sense, close to the Moses state under consideration. The correlations of simple operators will be dominated by contributions from intermediate states with only a few additional particles and/or holes near the generalized Fermi momenta, similar to the ground state case. It is a fact that the logic of bosonization is not strictly limited to ground states: one can explore the vicinity (in Hilbert space) of any zero-entropy state with finite velocities using the Tomonaga-Luttinger effective Hamiltonian logic. For Moses states,  
we  therefore  linearize the dispersion relation around the four points $k_{ia}$.  Indeed, it turns out that the situation is well described by a multicomponent Tomonaga-Luttinger model similar to that used in equilibrium cases \cite{1993_Penc_PRB_47,1995_Voit_RPP_58,2004_Pletyukhov_PRB_70}. As we will show, asymptotes of correlation functions are accurately reproduced, including exponents and prefactors, using the same technology as for ground state ones.

We approach the problem from the Tonks-Girardeau limit, recasting it in terms of fermions, and we project the fermion annihilation operator around the extremal momenta as
\begin{equation}
\label{eq:psi_f}
\Psi_F(x) \approx \sum_{ia}e^{i k_{ia} x} \psi_{ia}(x).
\end{equation}
Here, $\psi_{ia}(x)$ are chiral fermions with non-zero momentum modes in a restricted interval around zero.

We may now use the bosonization identity for the four chiral fermionic fields 
\begin{equation}
	\psi_{ia}(x) = \frac{1}{\sqrt{L}}   e^{-i\phi_{ia}(x)},
\end{equation}
where Klein factors and normal ordering are implicit.
The fields $\phi_{ia}(x)$ satisfy commutation relations
\begin{equation}
	[\phi_{ia}(x),\nabla \phi_{jb}(y)]=- s_a 2\pi i   \delta(x-y) \delta_{ia,jb},
\end{equation} 
where $\delta_{ia,jb}$ is the Kronecker delta and $s_R=1,\, s_L=-1$.
The bosonic fields relate to the density operators of the chiral fermions as 
\begin{equation}\label{eq:chiraldensityop}
	\rho_{ia}(x) =\psi_{ia}^\dagger(x) \psi_{ia}(x)= \frac{-s_a}{2\pi} \nabla \phi_{ia}(x).
\end{equation}
By linearizing the dispersion relation we obtain an effective Hamiltonian. In the Tonks-Girardeau limit, this may be written in terms of the bosonic fields as
\begin{equation}
	H^{TL}_0 = \sum_{ia} \frac{s_av_{ia} }{4\pi}\int dx (\nabla \phi_{ia}(x))^2.
\end{equation}
Here, the velocities $v_{ia} $ are like the Fermi velocities derived from  the dispersion relation  of the Tonks-Girardeau gas at the respective edges of the two seas. We include the sign factors $s_{a}$ to compensate for the ``wrong'' direction of the derivative for left edges, such that the  energy of excitations with small momenta on top of the Moses state have the right sign. Note however  that excitations in the region between the two seas have negative energies: $v_{1R},-v_{2L}<0$. 

To get away from the Tonks-Girardeau limit, we add density-density interactions between the chiral fermions
\begin{equation}
H^{TL}_{int}=\sum_{ia,jb}\int dx g_{ia,jb}  \rho_{ia}(x) \rho_{jb}(x).
\end{equation}
The Hamiltonian can be rediagonalized by a canonical transformation
such that it becomes a free boson Hamiltonian  again
\begin{equation}
H^{TL}_{Moses} =  \sum_{kc}  \frac{s_c\tilde{v}_{kc} }{4\pi}\int dx (\nabla \tilde{\phi}_{kc}(x))^2,
\end{equation}
with renormalized velocities $\tilde{v}_{ia}$. The behaviour of the rescaled velocities $\tilde{v}_{ia}$ as a function of $c$ for the Moses state is plotted in Fig.~\ref{fig:fermispeed_rapidities}. Again the sign $s_{R/L} = \pm 1$ implements the correct energy for left-movers for which velocities are measured to the right so that negative velocity corresponds to positive energy. The interaction is encoded  in the definition of the  free fields according  to
\begin{equation}
\phi_{ia}(x) = \sum_{kc} U_{ia,kc} \tilde{\phi}_{kc}(x).
\end{equation}
The free field correlator that we will frequently use is
\begin{equation}
	\langle e^{i \alpha\tilde{\phi}_{kc}(x)} e^{-i\alpha \tilde{\phi}_{kc}(0)}\rangle = \left( \frac{i s_c}{\rho_0 x}\right)^{\alpha^2}.
\end{equation}

Let us make the connection with the conventional Luttinger liquid. There we have only one Fermi sea and $k_{R/L}=\pm k_F$. The real-valued $U$ matrix is then related to the Luttinger parameter $K$ via
\begin{equation}\label{eq:Ulutliq}
	U= \begin{pmatrix}
		 	\frac{1}{2\sqrt{K}} +\frac{1}{2}\sqrt{K} &		  \frac{1}{2\sqrt{K}}-\frac{1}{2}\sqrt{K}\\
			\frac{1}{2\sqrt{K}}-\frac{1}{2}\sqrt{K} &		\frac{1}{2\sqrt{K}}+\frac{1}{2}\sqrt{K} 
	\end{pmatrix}.
\end{equation}
In our more general context, $U$ arises from a $(d>2)$-dimensional Bogoliubov transformation and is no longer parametrized by a single `Luttinger parameter', hence we keep its matrix elements explicit in our formulas. In order to respect the commutation relations of the bosonic fields it must satisfy the quasi-unitarity condition $(U^{-1})_{ia,jb} = s_a s_b U_{jb,ia}$.
It is well-known that the Luttinger parameter $K$ can  be obtained from the compressibility of the system in the ground state, which is easily obtained numerically by finite size calculations \cite{GiamarchiBOOK}. Similarly, $U$ can be obtained from finite size computations from the $1/L$ corrections to the spectrum, which are given by 
\begin{equation}
	H^{TL}_{1/L} = \frac{\pi}{L} s_as_bs_c U_{ia,kc} U_{jb,kc} \tilde{v}_{kc} \hat{N}_{ia} \hat{N}_{jb}.
\end{equation}
Here $\hat{N}_{ia}$ measures the excess number of particles on the $ia$ branch on top of the Moses state. In the case of symmetric seas this relates simply to the response to a change of total-particle number or relative filling of the seas \cite{1985_deVega_NPB_251,*1989_Woynarovich_JPA_22}, which in an equilibrium context can be obtained by the variation of the chemical potential and the effective magnetic field. In our case, the calculation proceeds as follows: for a given interaction value $c$, and for a chosen Moses state (in terms of a quantum number configuration), energies of states with various $N_{i,a}$ are computed by solving the relevant Bethe equations, and the $U_{ia, kc}$ are read off. The effective theory for states in the vicinity of the Moses state, which will be used later for fitting the correlations from integrability, is thus completely specified using energy data only. 

\section{Criticality in asymptotics of correlations}
\subsection{Density-density correlation}
\begin{figure}\centering
 \includegraphics[width=8.5cm]{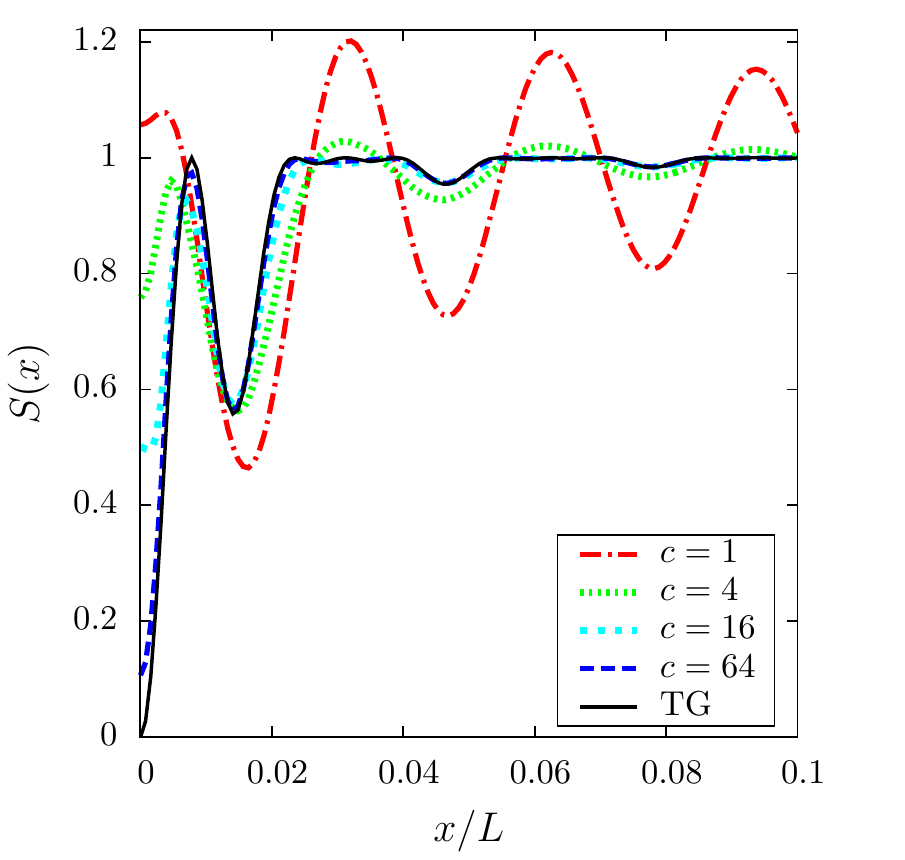}
 \caption{(Color online.) The density-density correlation function $S(x)$ for a Moses state for different values of the interaction obtained from the numerical data of the DSF as shown in Fig. \ref{fig:dsf}.}
 \label{fig:g2_changing_c_1}
\end{figure}

\begin{figure}\centering
 \includegraphics[width=8.5cm]{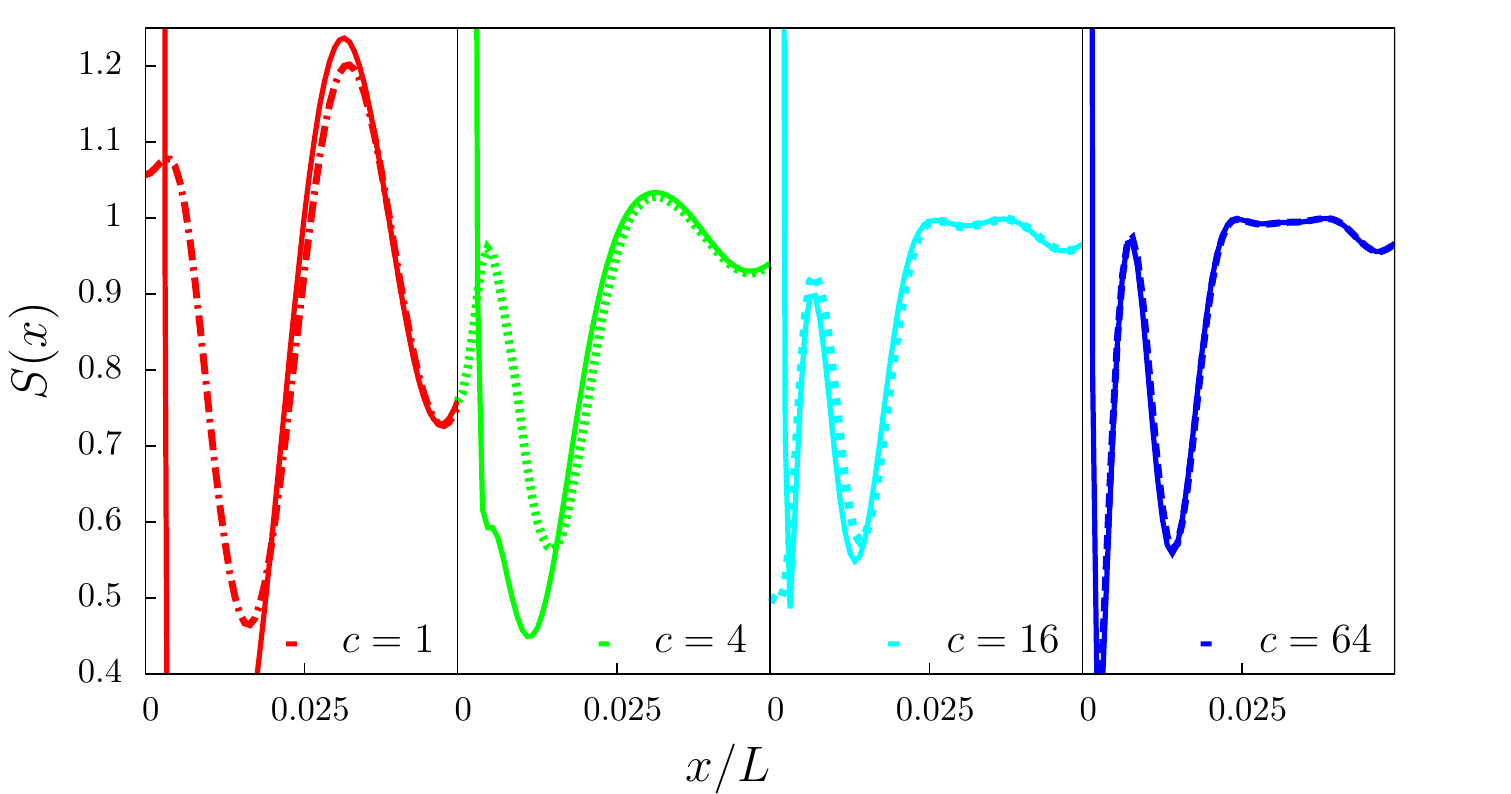}
 \caption{(Color online.) Comparison of the density-density correlation obtained numerically (dashed lines) and the analytic results from the multicomponent Tomonaga-Luttinger model (solid lines). The Tomonaga-Luttinger prediction differs from the numerical data by  less than $0.01$ when  $x$ is larger that $6.1\%$ of the system length for $c=1$, $2.8\%$ for $c=4$, $2.3\%$ for $c=16$  and $1.4\%$ for $c=64$. Calculations were performed at unit filling  with $64$ particles on length $L=64$.}
 \label{fig:g2_changing_c_2}
\end{figure}

The physical density operator is expressed in terms of chiral fermions as
\begin{equation}
\label{eq:dens}
	\rho(x)=\rho_0 + \sum_{ia} \rho_{ia}(x) +\sum_{ia\neq jb} e^{-i(k_{ia} - k_{jb})x} \psi^\dagger_{ia}(x)\psi_{jb}(x).\!
\end{equation}
The density-density correlation function 
\begin{equation}
	S(x) = \frac{\langle\rho(x)\rho(0)\rangle}{\rho_0^2} 
\end{equation}
 can easily be be obtained from the multicomponent Tomonaga-Luttinger model as
\begin{align}
	&S(x) = 1 -\frac{\sum_{ia,jb,kc}s_as_b U_{ia,kc}U_{jb,kc}}{4\pi^2 (\rho_0 x)^2} \\
	&+ \sum_{ia\neq jb} \frac{A_{ia,jb}}{4\pi^2}(-1)^{\delta_{ab}(1-\delta_{ij})} \cos((k_{ia}-k_{jb})x) \left( \frac{ 1}{\rho_0 x}\right)^{\mu_{ia,jb}} \nonumber
\end{align}
with
\begin{equation}
	\mu_{ia,jb} = \sum_{kc}(s_a U_{ia,kc} - s_b U_{jb,kc})^2.
\end{equation}
In the Tonks-Girardeau limit, $U_{ia,jb} = \delta_{ia,jb}$ and  $A_{ia,jb}=1$, the above expression reduces to the exact result  for any configuration of the Fermi seas with  edges $\{k_{1L}, k_{1R}, k_{2L}, k_{2R}\}$ as can be confirmed by an exact calculation.

\begin{figure}
 \includegraphics[width=8.5cm]{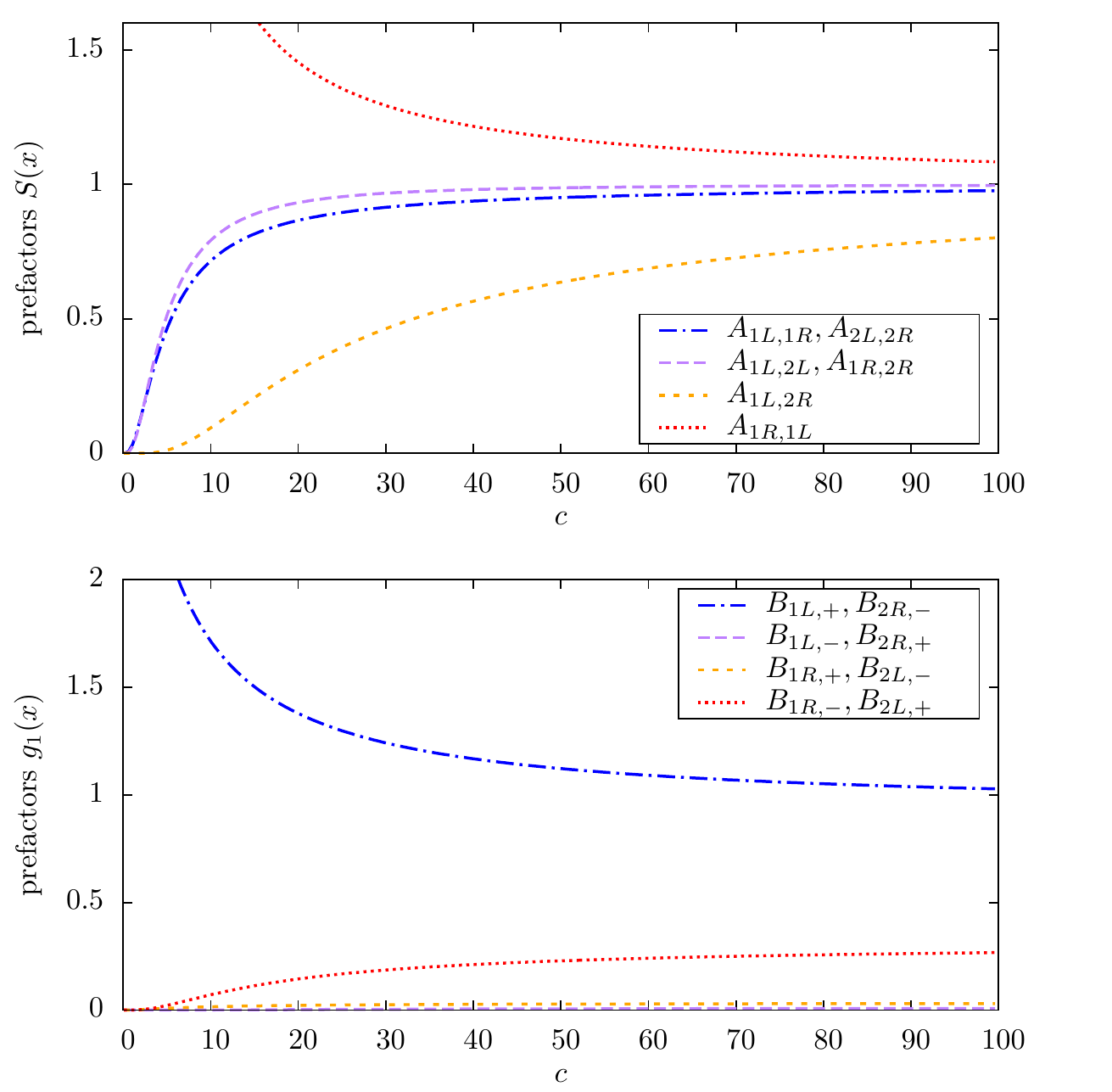}
\caption{(Color online.) Prefactors of the Tomonaga-Luttinger correlation asymptotics as a function of $c$.}
\label{fig:prefactors}
\end{figure}

The $A_{ia,jb}$ are nonuniversal prefactors, giving the  amplitude of the fluctuating terms (corresponding to Umklapp-like excitations). It was recently shown that the nonuniversal prefactors in Luttinger liquid correlations can be obtained from the finite size scaling of matrix elements \cite{2011_Shashi_PRB_84,2012_Shashi_PRB_85}. This logic can be carried over to the present context. In leading order, the matrix elements satisfy the scaling relation
\begin{equation}
 {|\langle ia,jb| \rho |M\rangle|^2 \over \rho_0^2}=  {A_{ia,jb} \over 4\pi^2}\left( {2\pi \over \rho_0 L }\right)^{\mu_{ia,jb}}
\end{equation}
where $|M\rangle$ denotes the Moses state and $|ia,jb\rangle$ the state obtained after creating an `Umklapp' excitation transferring a particle from the  $ia$ to the $jb$ branch or {\it vice versa} (see \cite{2011_Shashi_PRB_84,2012_Shashi_PRB_85} for the detailed explanations).  We obtained the scaling numerically by explicitly evaluating the relevant matrix elements for increasing system size. This provides the prefactors, which combined with the effective parameters of the Tomonaga-Luttinger model obtained above, yield a completely parameter-free fit for the correlations away from the Tonks-Girardeau regime. In addition, the exponents are obtained efficiently from the scaling of the prefactors, providing an independent check on the parameters determined from finite size corrections to the spectrum or a different route to obtaining the correlation exponents. In  Figure \ref{fig:g2_changing_c_1} the density-density correlation for different values of $c$ as obtained from the DSF presented in Figure~\ref{fig:dsf} is shown. In Figure~\ref{fig:g2_changing_c_2} we compare the field theory results with the numerical data. To fit the finite size data, we make the substitution $\rho_0 x \to {\pi\over \rho_0L} \sin(\pi x/L)$. There is excellent agreement for all distances larger than a fraction of the system length (Fig.~\ref{fig:g2_changing_c_2}). Figure.~\ref{fig:prefactors} shows the prefactors as a function of the interaction $c$. 

\subsection{The one-body density matrix}
The one-body reduced density matrix is given by
\begin{equation}
\begin{split}
&g_1(x)={\langle \Psi^\dagger (x) \Psi(0) \rangle\over \rho_0} \\
\end{split}
\label{eq:1pdensitymatrix}
\end{equation}
which is simply the Fourier transform of the momentum distribution functions.

\begin{figure}
 \includegraphics[width=8.5cm]{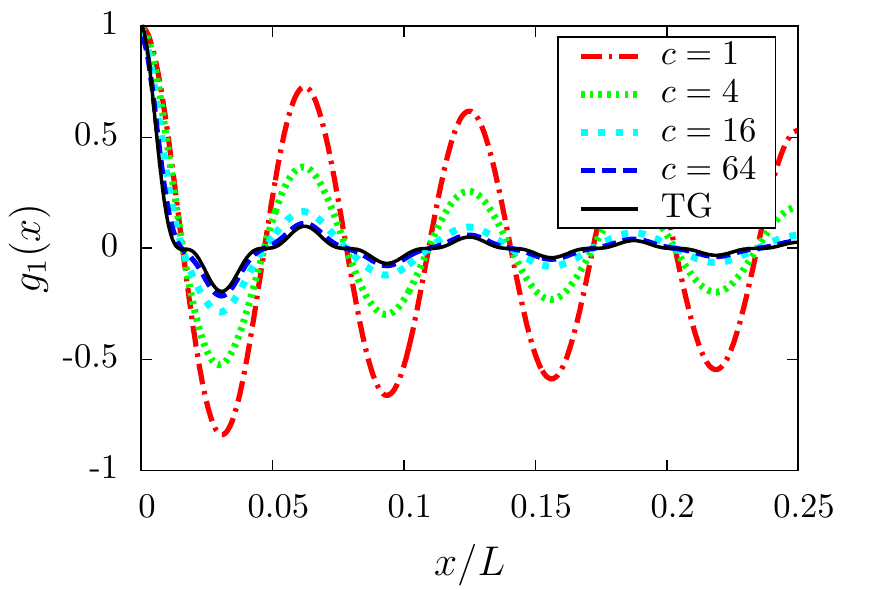}
\caption{(Color online.) The one-body reduced density matrix $g_1(x)$ for a Moses state for different values of the interaction as obtained from the numerical data of the momentum distribution function in Fig. \ref{fig:n_k}.}
\label{fig:g1_1}
\end{figure}

\begin{figure}
 \includegraphics[width=8.5cm]{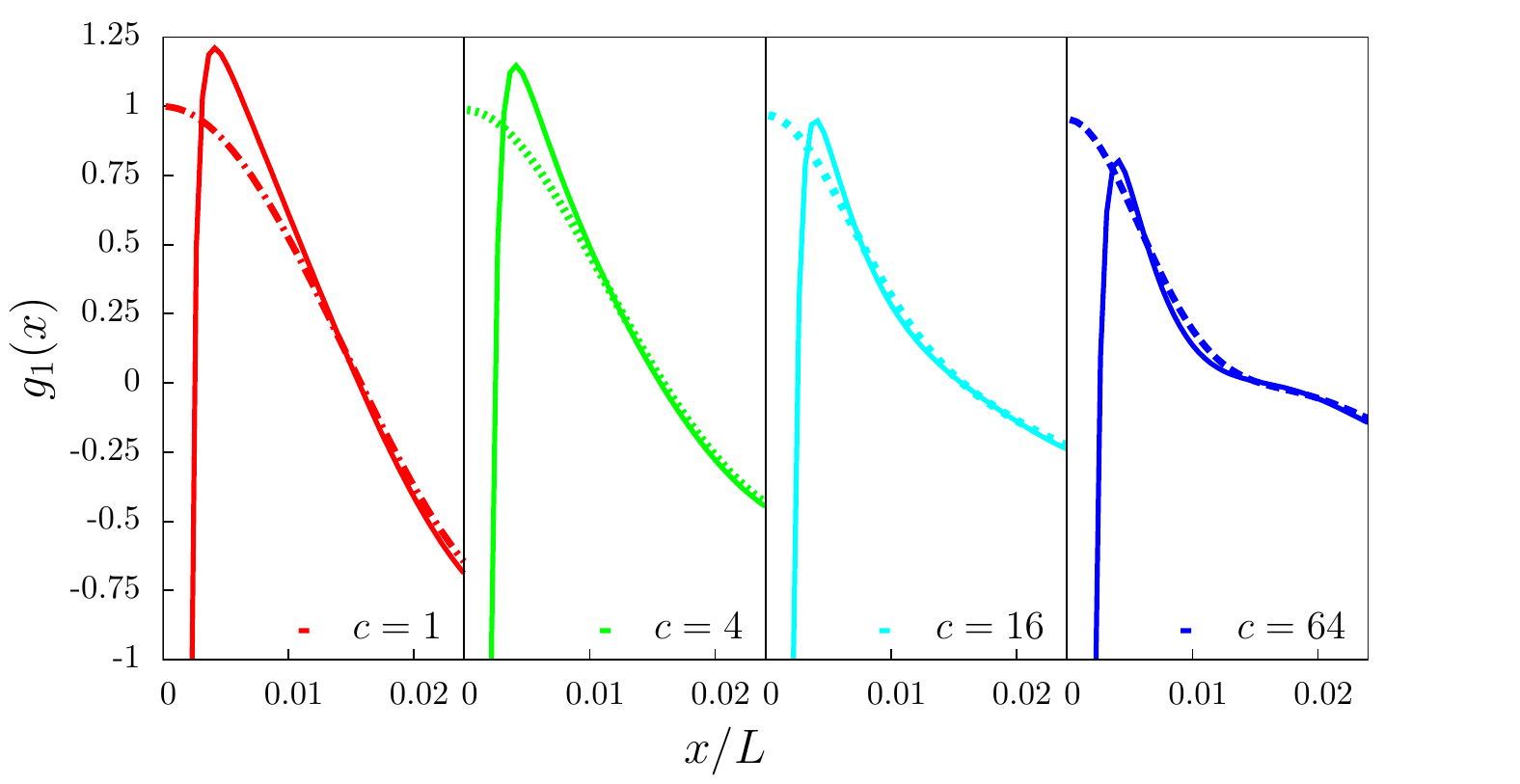}
\caption{(Color online.) Comparison of the one-body reduced density matrix obtained numerically (dashed lines) with the parameter free fits using the multicomponent Tomonaga-Luttinger model (solid lines). The difference between the numerical plot and the analytic multicomponent Tomonaga-Luttinger curve becomes less then $0.01$ for $x$ greater than $0.038 L$, for $c=1$ $ 0.028L$ for $c=4$, $0.030 L$ for $c=16$ and $c=64$.}
\label{fig:g1_2}
\end{figure}

In order to obtain $g_1(x)$ in our Tomonaga-Luttinger description one must be careful to take particle statistics into account. We therefore introduce a Jordan-Wigner string operator and define the boson annihilation operator as
\begin{equation}
	\Psi(x) \equiv \cos\left(\pi \int_0^x dy \rho(y) \right)\Psi_F(x),
\end{equation}
where $\Psi_F(x)$ has been defined in (\ref{eq:psi_f}).
The physical density operator is given in equation \eqref{eq:dens}. Neglecting the fast fluctuating terms of the density operator under the integral, and using (\ref{eq:chiraldensityop}) we find the expression
\begin{equation}
	\int_0^x dy \rho(y) = \frac{k_F}{\pi}x-\sum_{i,a} \frac{s_a }{2\pi} \phi_{ia}(x),
\end{equation}
leading to
\begin{equation}
\begin{split}
	\Psi(x) \approx\ \frac{1}{2\sqrt{L}}\sum_{ia}\sum_{\epsilon =\pm 1} e^{i(k_{ia} + \epsilon k_F)x}\\
	\times e^{ -i \sum_{kc} (\epsilon s_c /2 + \delta_{ia,kc})\phi_{kc}(x)}.
\end{split}
\end{equation}
The one-body function is readily computed as
\begin{equation}
  g_1(x) = 
 \sum_{ia,\epsilon}{B_{ia,\epsilon} \over 2\pi} (-1)^{\delta_{s_a, \epsilon}} e^{-i k_{ia,\epsilon}x}
	 \left(\frac{1}{\rho_0 x}\right)^{\mu_{ia,\epsilon}}.
\end{equation}
where
\begin{equation}
	\mu_{ia,\epsilon}=\sum_{ld} \left[\sum_{kc} ( \epsilon/2 +s_a\delta_{ia,ld})U_{kc,ld}\right]^2,
\end{equation}
with the notation $k_{ia,\epsilon}\equiv k_{ia} +\epsilon k_F$. The non-universal prefactors $B_{ia,\epsilon}$ that have to be obtained independently from
\begin{equation}
{|\langle ia,\epsilon | \Psi |M\rangle|^2\over \rho_0 } =  {B_{ia,\epsilon}\over 2\pi} \left( {2\pi \over \rho_0 L }\right)^{\mu_{ia,\epsilon}},
\end{equation}
according to a procedure similar to that described above for the DSF (see also Fig.~\ref{fig:prefactors}).
The correlation is shown for different values of $c$ in Fig.~\ref{fig:g1_1} and Fig.~\ref{fig:g1_2}.

The sign $\epsilon$ in the Jordan-Wigner string operator shifts the momenta $k_{ia}$ by $k_F$ to either the left or right. It absorbs the mismatch of the quantum number lattices in the Bethe Ansatz solution for even and odd numbers of particles: it corresponds to the choice of moving all occupied quantum numbers by a half to the left or to the right after removing a single particle from the system.  The  prefactors (see Fig. \ref{fig:prefactors}) and the exponents both show the relative importance of the contributions with $\epsilon s_a = -1$: these have a much larger contribution and decay more slowly. Indeed, this clarifies the position of the  peaks at $\{k_{1L}+k_F,k_{1R}-k_F, k_{2L}+k_F,k_{2R}-k_F\}$ in the momentum distribution function that were mentioned above. 

From the expression for $g_1(x)$ we find for small $k$ around $k_{ia,\epsilon}$ the result
\begin{equation}
	n(k-k_{ia,\epsilon}) \sim |k-k_{ia,\epsilon}|^{\mu_{ia,\epsilon}-1}.
\end{equation}
 In the limit of infinite repulsion this becomes
\begin{equation}\label{eq:muTG}
\mu^{c=\infty}_{ia,\epsilon} = 1+ \frac{n}{2}+ \epsilon s_a,
\end{equation}
where $n$ is the total number of seas.

The power law at zero momentum for a gas of bosons in the ground state is obtained from reduction to the conventional  Luttinger liquid, i.e. with $U$ given by Eq.~\eqref{eq:Ulutliq}. This leads to  the well-known result \cite{2004_Cazalilla_JPB_37}
\begin{equation}
	n(k) \sim k^{\mu_0-1} \quad\text{with}\quad \mu_0 =\frac{1}{2K}.
\end{equation}
 Choosing $s_{ia} = 1,\epsilon=-1$ in  Eq.~\eqref{eq:muTG} indeed gives the correct result $\mu_0^{c=\infty} = 1/2$ for the Tonks-Girardeau ground state ($K=1$). 

At large momenta, the MDF decays as $1/k^4$ as expected from the logic of Tan's contact \cite{2008_Tan_AP_323_1}. We have also directly verified this from the small-$x$ expansion of $g_1(x)$ in the Tonks-Girardeau limit.

\section{Conclusions and outlook}
We have studied a particular class of highly-excited states in the Lieb-Liniger model, obtained by splitting the ground state Fermi sea, giving finite different momentum to macroscopic subsets of atoms. These `Moses' states possess a richer excitation spectrum than the ground state, and display a number of interesting features in their correlations, namely extra branches, critical power-law like behaviour and nontrivial threshold exponents. We have shown that the integrability-based results obtained could be very well fitted using a multicomponent Tomonaga-Luttinger description, whose effective parameters are set by energy (and thus Bethe Ansatz-obtainable) data. The threshold behavior can be studied in more detail (explicitly giving the interaction and momentum-dependent threshold exponents) by adapting methods from nonlinear Luttinger liquid theory \cite{2008_Imambekov_PRL_100,2009_Imambekov_SCIENCE_323}. It is completely straightforward to generalize our results to the case of multiple seas, although the computation of correlations becomes increasingly difficult.

One interesting generalization is to consider `thermal-like' dressing of Moses states. The ensemble of states thus obtained could be used to model the initial state immediately after a Bragg pulse, as performed in the quantum Newton's cradle experiment of \cite{2006_Kinoshita_NATURE_440}. We will investigate this issue in future publications.

\section*{Acknowledgements}
\begin{acknowledgments}
We would like to thank  M. Brockmann, V. Gritsev, J. Mossel, J. De Nardis, M. Panfil, B. Wouters, M. Rigol and D. S. Weiss for useful discussions.  We acknowledge support from the Foundation for Fundamental Research on Matter (FOM) and from the Netherlands Organisation for Scientific Research (NWO).
\end{acknowledgments}


\begin{thebibliography}{43}%
\makeatletter
\providecommand \@ifxundefined [1]{%
 \@ifx{#1\undefined}
}%
\providecommand \@ifnum [1]{%
 \ifnum #1\expandafter \@firstoftwo
 \else \expandafter \@secondoftwo
 \fi
}%
\providecommand \@ifx [1]{%
 \ifx #1\expandafter \@firstoftwo
 \else \expandafter \@secondoftwo
 \fi
}%
\providecommand \natexlab [1]{#1}%
\providecommand \enquote  [1]{``#1''}%
\providecommand \bibnamefont  [1]{#1}%
\providecommand \bibfnamefont [1]{#1}%
\providecommand \citenamefont [1]{#1}%
\providecommand \href@noop [0]{\@secondoftwo}%
\providecommand \href [0]{\begingroup \@sanitize@url \@href}%
\providecommand \@href[1]{\@@startlink{#1}\@@href}%
\providecommand \@@href[1]{\endgroup#1\@@endlink}%
\providecommand \@sanitize@url [0]{\catcode `\\12\catcode `\$12\catcode
  `\&12\catcode `\#12\catcode `\^12\catcode `\_12\catcode `\%12\relax}%
\providecommand \@@startlink[1]{}%
\providecommand \@@endlink[0]{}%
\providecommand \url  [0]{\begingroup\@sanitize@url \@url }%
\providecommand \@url [1]{\endgroup\@href {#1}{\urlprefix }}%
\providecommand \urlprefix  [0]{URL }%
\providecommand \Eprint [0]{\href }%
\providecommand \doibase [0]{http://dx.doi.org/}%
\providecommand \selectlanguage [0]{\@gobble}%
\providecommand \bibinfo  [0]{\@secondoftwo}%
\providecommand \bibfield  [0]{\@secondoftwo}%
\providecommand \translation [1]{[#1]}%
\providecommand \BibitemOpen [0]{}%
\providecommand \bibitemStop [0]{}%
\providecommand \bibitemNoStop [0]{.\EOS\space}%
\providecommand \EOS [0]{\spacefactor3000\relax}%
\providecommand \BibitemShut  [1]{\csname bibitem#1\endcsname}%
\let\auto@bib@innerbib\@empty
\bibitem [{\citenamefont {Giamarchi}(2004)}]{GiamarchiBOOK}%
  \BibitemOpen
  \bibfield  {author} {\bibinfo {author} {\bibfnamefont {T.}~\bibnamefont
  {Giamarchi}},\ }\href@noop {} {\emph {\bibinfo {title} {Quantum Physics in
  One Dimension}}}\ (\bibinfo  {publisher} {Oxford University Press},\ \bibinfo
  {year} {2004})\BibitemShut {NoStop}%
\bibitem [{\citenamefont {Haldane}(1981{\natexlab{a}})}]{1981_Haldane_PRL_47}%
  \BibitemOpen
  \bibfield  {author} {\bibinfo {author} {\bibfnamefont {F.~D.~M.}\
  \bibnamefont {Haldane}},\ }\href {\doibase 10.1103/PhysRevLett.47.1840}
  {\bibfield  {journal} {\bibinfo  {journal} {Phys. Rev. Lett.}\ }\textbf
  {\bibinfo {volume} {47}},\ \bibinfo {pages} {1840} (\bibinfo {year}
  {1981}{\natexlab{a}})}\BibitemShut {NoStop}%
\bibitem [{\citenamefont {Haldane}(1981{\natexlab{b}})}]{1981_Haldane_JPC_14}%
  \BibitemOpen
  \bibfield  {author} {\bibinfo {author} {\bibfnamefont {F.~D.~M.}\
  \bibnamefont {Haldane}},\ }\href@noop {} {\bibfield  {journal} {\bibinfo
  {journal} {J. Phys C: Sol. St. Phys.}\ }\textbf {\bibinfo {volume} {14}},\
  \bibinfo {pages} {2585} (\bibinfo {year} {1981}{\natexlab{b}})}\BibitemShut
  {NoStop}%
\bibitem [{\citenamefont {Gogolin}\ \emph {et~al.}(1998)\citenamefont
  {Gogolin}, \citenamefont {Nersesyan},\ and\ \citenamefont
  {Tsvelik}}]{GogolinBOOK}%
  \BibitemOpen
  \bibfield  {author} {\bibinfo {author} {\bibfnamefont {A.~O.}\ \bibnamefont
  {Gogolin}}, \bibinfo {author} {\bibfnamefont {A.~A.}\ \bibnamefont
  {Nersesyan}}, \ and\ \bibinfo {author} {\bibfnamefont {A.~M.}\ \bibnamefont
  {Tsvelik}},\ }\href@noop {} {\emph {\bibinfo {title} {{Bosonization and
  Strongly Correlated Systems}}}}\ (\bibinfo  {publisher} {Cambridge University
  Press},\ \bibinfo {year} {1998})\BibitemShut {NoStop}%
\bibitem [{\citenamefont {von Delft}\ and\ \citenamefont
  {Schoeller}(1998)}]{1998_von_Delft_AP_7}%
  \BibitemOpen
  \bibfield  {author} {\bibinfo {author} {\bibfnamefont {J.}~\bibnamefont {von
  Delft}}\ and\ \bibinfo {author} {\bibfnamefont {H.}~\bibnamefont
  {Schoeller}},\ }\href@noop {} {\bibfield  {journal} {\bibinfo  {journal}
  {Ann. Phys. (Leipzig)}\ }\textbf {\bibinfo {volume} {7}},\ \bibinfo {pages}
  {225} (\bibinfo {year} {1998})}\BibitemShut {NoStop}%
\bibitem [{\citenamefont {Korepin}\ \emph {et~al.}(1993)\citenamefont
  {Korepin}, \citenamefont {Bogoliubov},\ and\ \citenamefont
  {Izergin}}]{KorepinBOOK}%
  \BibitemOpen
  \bibfield  {author} {\bibinfo {author} {\bibfnamefont {V.~E.}\ \bibnamefont
  {Korepin}}, \bibinfo {author} {\bibfnamefont {N.~M.}\ \bibnamefont
  {Bogoliubov}}, \ and\ \bibinfo {author} {\bibfnamefont {A.~G.}\ \bibnamefont
  {Izergin}},\ }\href@noop {} {\emph {\bibinfo {title} {Quantum Inverse
  Scattering Method and Correlation Functions}}}\ (\bibinfo  {publisher}
  {Cambridge Univ. Press},\ \bibinfo {year} {1993})\BibitemShut {NoStop}%
\bibitem [{\citenamefont {Bethe}(1931)}]{1931_Bethe_ZP_71}%
  \BibitemOpen
  \bibfield  {author} {\bibinfo {author} {\bibfnamefont {H.}~\bibnamefont
  {Bethe}},\ }\href@noop {} {\bibfield  {journal} {\bibinfo  {journal} {Zeit.
  f\"ur Physik}\ }\textbf {\bibinfo {volume} {71}},\ \bibinfo {pages} {205}
  (\bibinfo {year} {1931})}\BibitemShut {NoStop}%
\bibitem [{\citenamefont {Panfil}\ \emph {et~al.}(2013)\citenamefont {Panfil},
  \citenamefont {De~Nardis},\ and\ \citenamefont {Caux}}]{2013_Panfil_PRL_110}%
  \BibitemOpen
  \bibfield  {author} {\bibinfo {author} {\bibfnamefont {M.}~\bibnamefont
  {Panfil}}, \bibinfo {author} {\bibfnamefont {J.}~\bibnamefont {De~Nardis}}, \
  and\ \bibinfo {author} {\bibfnamefont {J.-S.}\ \bibnamefont {Caux}},\
  }\href@noop {} {\bibfield  {journal} {\bibinfo  {journal} {Phys. Rev. Lett.}\
  }\textbf {\bibinfo {volume} {110}},\ \bibinfo {pages} {125302} (\bibinfo
  {year} {2013})}\BibitemShut {NoStop}%
\bibitem [{\citenamefont {Bloch}\ \emph {et~al.}(2008)\citenamefont {Bloch},
  \citenamefont {Dalibard},\ and\ \citenamefont {Zwerger}}]{2008_Bloch_RMP_80}%
  \BibitemOpen
  \bibfield  {author} {\bibinfo {author} {\bibfnamefont {I.}~\bibnamefont
  {Bloch}}, \bibinfo {author} {\bibfnamefont {J.}~\bibnamefont {Dalibard}}, \
  and\ \bibinfo {author} {\bibfnamefont {W.}~\bibnamefont {Zwerger}},\ }\href
  {\doibase 10.1103/RevModPhys.80.885} {\bibfield  {journal} {\bibinfo
  {journal} {Rev. Mod. Phys.}\ }\textbf {\bibinfo {volume} {80}},\ \bibinfo
  {pages} {885} (\bibinfo {year} {2008})}\BibitemShut {NoStop}%
\bibitem [{\citenamefont {Cazalilla}\ \emph {et~al.}(2011)\citenamefont
  {Cazalilla}, \citenamefont {Citro}, \citenamefont {Giamarchi}, \citenamefont
  {Orignac},\ and\ \citenamefont {Rigol}}]{2011_Cazalilla_RMP_83}%
  \BibitemOpen
  \bibfield  {author} {\bibinfo {author} {\bibfnamefont {M.~A.}\ \bibnamefont
  {Cazalilla}}, \bibinfo {author} {\bibfnamefont {R.}~\bibnamefont {Citro}},
  \bibinfo {author} {\bibfnamefont {T.}~\bibnamefont {Giamarchi}}, \bibinfo
  {author} {\bibfnamefont {E.}~\bibnamefont {Orignac}}, \ and\ \bibinfo
  {author} {\bibfnamefont {M.}~\bibnamefont {Rigol}},\ }\href {\doibase
  10.1103/RevModPhys.83.1405} {\bibfield  {journal} {\bibinfo  {journal} {Rev.
  Mod. Phys.}\ }\textbf {\bibinfo {volume} {83}},\ \bibinfo {pages} {1405}
  (\bibinfo {year} {2011})}\BibitemShut {NoStop}%
\bibitem [{\citenamefont {Kinoshita}\ \emph {et~al.}(2006)\citenamefont
  {Kinoshita}, \citenamefont {Wenger},\ and\ \citenamefont
  {Weiss}}]{2006_Kinoshita_NATURE_440}%
  \BibitemOpen
  \bibfield  {author} {\bibinfo {author} {\bibfnamefont {T.}~\bibnamefont
  {Kinoshita}}, \bibinfo {author} {\bibfnamefont {T.}~\bibnamefont {Wenger}}, \
  and\ \bibinfo {author} {\bibfnamefont {D.~S.}\ \bibnamefont {Weiss}},\
  }\href@noop {} {\bibfield  {journal} {\bibinfo  {journal} {Nature}\ }\textbf
  {\bibinfo {volume} {440}},\ \bibinfo {pages} {900} (\bibinfo {year}
  {2006})}\BibitemShut {NoStop}%
\bibitem [{\citenamefont {Lieb}\ and\ \citenamefont
  {Liniger}(1963)}]{1963_Lieb_PR_130_1}%
  \BibitemOpen
  \bibfield  {author} {\bibinfo {author} {\bibfnamefont {E.~H.}\ \bibnamefont
  {Lieb}}\ and\ \bibinfo {author} {\bibfnamefont {W.}~\bibnamefont {Liniger}},\
  }\href {\doibase 10.1103/PhysRev.130.1605} {\bibfield  {journal} {\bibinfo
  {journal} {Phys. Rev.}\ }\textbf {\bibinfo {volume} {130}},\ \bibinfo {pages}
  {1605} (\bibinfo {year} {1963})}\BibitemShut {NoStop}%
\bibitem [{\citenamefont {Lieb}(1963)}]{1963_Lieb_PR_130_2}%
  \BibitemOpen
  \bibfield  {author} {\bibinfo {author} {\bibfnamefont {E.~H.}\ \bibnamefont
  {Lieb}},\ }\href {\doibase 10.1103/PhysRev.130.1616} {\bibfield  {journal}
  {\bibinfo  {journal} {Phys. Rev.}\ }\textbf {\bibinfo {volume} {130}},\
  \bibinfo {pages} {1616} (\bibinfo {year} {1963})}\BibitemShut {NoStop}%
\bibitem [{\citenamefont {Girardeau}(1960)}]{1960_Girardeau_JMP_1}%
  \BibitemOpen
  \bibfield  {author} {\bibinfo {author} {\bibfnamefont {M.}~\bibnamefont
  {Girardeau}},\ }\href {\doibase 10.1063/1.1703687} {\bibfield  {journal}
  {\bibinfo  {journal} {J. Math. Phys.}\ }\textbf {\bibinfo {volume} {1}},\
  \bibinfo {pages} {516} (\bibinfo {year} {1960})}\BibitemShut {NoStop}%
\bibitem [{\citenamefont {Paredes}\ \emph {et~al.}(2004)\citenamefont
  {Paredes}, \citenamefont {Widera}, \citenamefont {Murg}, \citenamefont
  {Mandel}, \citenamefont {F{\"o}lling}, \citenamefont {Cirac}, \citenamefont
  {Shlyapnikov}, \citenamefont {H{\"a}nsch},\ and\ \citenamefont
  {Bloch}}]{2004_Paredes_NATURE_429}%
  \BibitemOpen
  \bibfield  {author} {\bibinfo {author} {\bibfnamefont {B.}~\bibnamefont
  {Paredes}}, \bibinfo {author} {\bibfnamefont {A.}~\bibnamefont {Widera}},
  \bibinfo {author} {\bibfnamefont {V.}~\bibnamefont {Murg}}, \bibinfo {author}
  {\bibfnamefont {O.}~\bibnamefont {Mandel}}, \bibinfo {author} {\bibfnamefont
  {S.}~\bibnamefont {F{\"o}lling}}, \bibinfo {author} {\bibfnamefont
  {I.}~\bibnamefont {Cirac}}, \bibinfo {author} {\bibfnamefont {G.~V.}\
  \bibnamefont {Shlyapnikov}}, \bibinfo {author} {\bibfnamefont {T.~W.}\
  \bibnamefont {H{\"a}nsch}}, \ and\ \bibinfo {author} {\bibfnamefont
  {I.}~\bibnamefont {Bloch}},\ }\href@noop {} {\bibfield  {journal} {\bibinfo
  {journal} {Nature}\ }\textbf {\bibinfo {volume} {429}},\ \bibinfo {pages}
  {277} (\bibinfo {year} {2004})}\BibitemShut {NoStop}%
\bibitem [{\citenamefont {Kormos}\ \emph
  {et~al.}(2011{\natexlab{a}})\citenamefont {Kormos}, \citenamefont {Chou},\
  and\ \citenamefont {Imambekov}}]{2011_Kormos_PRL_107}%
  \BibitemOpen
  \bibfield  {author} {\bibinfo {author} {\bibfnamefont {M.}~\bibnamefont
  {Kormos}}, \bibinfo {author} {\bibfnamefont {Y.-Z.}\ \bibnamefont {Chou}}, \
  and\ \bibinfo {author} {\bibfnamefont {A.}~\bibnamefont {Imambekov}},\ }\href
  {\doibase 10.1103/PhysRevLett.107.230405} {\bibfield  {journal} {\bibinfo
  {journal} {Phys. Rev. Lett.}\ }\textbf {\bibinfo {volume} {107}},\ \bibinfo
  {pages} {230405} (\bibinfo {year} {2011}{\natexlab{a}})}\BibitemShut
  {NoStop}%
\bibitem [{\citenamefont {Pozsgay}(2011)}]{2011_Pozsgay_JSTAT_P01011}%
  \BibitemOpen
  \bibfield  {author} {\bibinfo {author} {\bibfnamefont {B.}~\bibnamefont
  {Pozsgay}},\ }\href {http://stacks.iop.org/1742-5468/2011/i=01/a=P01011}
  {\bibfield  {journal} {\bibinfo  {journal} {J. Stat. Mech.: Th. Exp.}\
  }\textbf {\bibinfo {volume} {2011}},\ \bibinfo {pages} {P01011} (\bibinfo
  {year} {2011})}\BibitemShut {NoStop}%
\bibitem [{\citenamefont {Kormos}\ \emph
  {et~al.}(2011{\natexlab{b}})\citenamefont {Kormos}, \citenamefont
  {Mussardo},\ and\ \citenamefont {Trombettoni}}]{2011_Kormos_PRA_83}%
  \BibitemOpen
  \bibfield  {author} {\bibinfo {author} {\bibfnamefont {M.}~\bibnamefont
  {Kormos}}, \bibinfo {author} {\bibfnamefont {G.}~\bibnamefont {Mussardo}}, \
  and\ \bibinfo {author} {\bibfnamefont {A.}~\bibnamefont {Trombettoni}},\
  }\href {\doibase 10.1103/PhysRevA.83.013617} {\bibfield  {journal} {\bibinfo
  {journal} {Phys. Rev. A}\ }\textbf {\bibinfo {volume} {83}},\ \bibinfo
  {pages} {013617} (\bibinfo {year} {2011}{\natexlab{b}})}\BibitemShut
  {NoStop}%
\bibitem [{\citenamefont {Kormos}\ \emph {et~al.}(2009)\citenamefont {Kormos},
  \citenamefont {Mussardo},\ and\ \citenamefont
  {Trombettoni}}]{2009_Kormos_PRL_103}%
  \BibitemOpen
  \bibfield  {author} {\bibinfo {author} {\bibfnamefont {M.}~\bibnamefont
  {Kormos}}, \bibinfo {author} {\bibfnamefont {G.}~\bibnamefont {Mussardo}}, \
  and\ \bibinfo {author} {\bibfnamefont {A.}~\bibnamefont {Trombettoni}},\
  }\href {\doibase 10.1103/PhysRevLett.103.210404} {\bibfield  {journal}
  {\bibinfo  {journal} {Phys. Rev. Lett.}\ }\textbf {\bibinfo {volume} {103}},\
  \bibinfo {pages} {210404} (\bibinfo {year} {2009})}\BibitemShut {NoStop}%
\bibitem [{\citenamefont {Rigol}\ \emph {et~al.}(2008)\citenamefont {Rigol},
  \citenamefont {Dunjko},\ and\ \citenamefont
  {Olshanii}}]{2008_Rigol_NATURE_452}%
  \BibitemOpen
  \bibfield  {author} {\bibinfo {author} {\bibfnamefont {M.}~\bibnamefont
  {Rigol}}, \bibinfo {author} {\bibfnamefont {V.}~\bibnamefont {Dunjko}}, \
  and\ \bibinfo {author} {\bibfnamefont {M.}~\bibnamefont {Olshanii}},\
  }\href@noop {} {\bibfield  {journal} {\bibinfo  {journal} {Nature}\ }\textbf
  {\bibinfo {volume} {452}},\ \bibinfo {pages} {854} (\bibinfo {year}
  {2008})}\BibitemShut {NoStop}%
\bibitem [{\citenamefont {Eriksson}\ and\ \citenamefont
  {Korepin}(2013)}]{2013_Eriksson_JPA_46}%
  \BibitemOpen
  \bibfield  {author} {\bibinfo {author} {\bibfnamefont {E.}~\bibnamefont
  {Eriksson}}\ and\ \bibinfo {author} {\bibfnamefont {V.}~\bibnamefont
  {Korepin}},\ }\href {http://stacks.iop.org/1751-8121/46/i=23/a=235002}
  {\bibfield  {journal} {\bibinfo  {journal} {J. Phys. A: Math. Theor.}\
  }\textbf {\bibinfo {volume} {46}},\ \bibinfo {pages} {235002} (\bibinfo
  {year} {2013})}\BibitemShut {NoStop}%
\bibitem [{\citenamefont {Frahm}\ and\ \citenamefont
  {R\"odenbeck}(1999)}]{Frahm:1999}%
  \BibitemOpen
  \bibfield  {author} {\bibinfo {author} {\bibfnamefont {H.}~\bibnamefont
  {Frahm}}\ and\ \bibinfo {author} {\bibfnamefont {C.}~\bibnamefont
  {R\"odenbeck}},\ }\href@noop {} {\bibfield  {journal} {\bibinfo  {journal}
  {Eur. Phys. J. B}\ }\textbf {\bibinfo {volume} {10}},\ \bibinfo {pages} {409}
  (\bibinfo {year} {1999})}\BibitemShut {NoStop}%
\bibitem [{\citenamefont {Frahm}\ and\ \citenamefont
  {R\"odenbeck}(1997)}]{1997_Frahm_JPA_30}%
  \BibitemOpen
  \bibfield  {author} {\bibinfo {author} {\bibfnamefont {H.}~\bibnamefont
  {Frahm}}\ and\ \bibinfo {author} {\bibfnamefont {C.}~\bibnamefont
  {R\"odenbeck}},\ }\href {http://stacks.iop.org/0305-4470/30/i=13/a=005}
  {\bibfield  {journal} {\bibinfo  {journal} {Journal of Physics A:
  Mathematical and General}\ }\textbf {\bibinfo {volume} {30}},\ \bibinfo
  {pages} {4467} (\bibinfo {year} {1997})}\BibitemShut {NoStop}%
\bibitem [{\citenamefont {Zvyagin}\ \emph {et~al.}(2001)\citenamefont
  {Zvyagin}, \citenamefont {Kl\"umper},\ and\ \citenamefont
  {Zittartz}}]{2001_Zvyagin_EPB_19}%
  \BibitemOpen
  \bibfield  {author} {\bibinfo {author} {\bibfnamefont {A.}~\bibnamefont
  {Zvyagin}}, \bibinfo {author} {\bibfnamefont {A.}~\bibnamefont {Kl\"umper}},
  \ and\ \bibinfo {author} {\bibfnamefont {J.}~\bibnamefont {Zittartz}},\
  }\href@noop {} {\bibfield  {journal} {\bibinfo  {journal} {Eur. Phys. J. B}\
  }\textbf {\bibinfo {volume} {19}},\ \bibinfo {pages} {25} (\bibinfo {year}
  {2001})}\BibitemShut {NoStop}%
\bibitem [{\citenamefont {Sato}\ \emph {et~al.}(2012)\citenamefont {Sato},
  \citenamefont {Kanamoto}, \citenamefont {Kaminishi},\ and\ \citenamefont
  {Deguchi}}]{2012_Sato_PRL_108}%
  \BibitemOpen
  \bibfield  {author} {\bibinfo {author} {\bibfnamefont {J.}~\bibnamefont
  {Sato}}, \bibinfo {author} {\bibfnamefont {R.}~\bibnamefont {Kanamoto}},
  \bibinfo {author} {\bibfnamefont {E.}~\bibnamefont {Kaminishi}}, \ and\
  \bibinfo {author} {\bibfnamefont {T.}~\bibnamefont {Deguchi}},\ }\href
  {\doibase 10.1103/PhysRevLett.108.110401} {\bibfield  {journal} {\bibinfo
  {journal} {Phys. Rev. Lett.}\ }\textbf {\bibinfo {volume} {108}},\ \bibinfo
  {pages} {110401} (\bibinfo {year} {2012})}\BibitemShut {NoStop}%
\bibitem [{\citenamefont {Caux}(2009)}]{2009_Caux_JMP_50}%
  \BibitemOpen
  \bibfield  {author} {\bibinfo {author} {\bibfnamefont {J.-S.}\ \bibnamefont
  {Caux}},\ }\href {\doibase 10.1063/1.3216474} {\bibfield  {journal} {\bibinfo
   {journal} {J. Math. Phys.}\ }\textbf {\bibinfo {volume} {50}},\ \bibinfo
  {eid} {095214} (\bibinfo {year} {2009})}\BibitemShut {NoStop}%
\bibitem [{\citenamefont {Caux}\ and\ \citenamefont
  {Calabrese}(2006)}]{2006_Caux_PRA_74}%
  \BibitemOpen
  \bibfield  {author} {\bibinfo {author} {\bibfnamefont {J.-S.}\ \bibnamefont
  {Caux}}\ and\ \bibinfo {author} {\bibfnamefont {P.}~\bibnamefont
  {Calabrese}},\ }\href {\doibase 10.1103/PhysRevA.74.031605} {\bibfield
  {journal} {\bibinfo  {journal} {Phys. Rev. A}\ }\textbf {\bibinfo {volume}
  {74}},\ \bibinfo {pages} {031605} (\bibinfo {year} {2006})}\BibitemShut
  {NoStop}%
\bibitem [{\citenamefont {Lenard}(1964)}]{1964_Lenard_JMP_5}%
  \BibitemOpen
  \bibfield  {author} {\bibinfo {author} {\bibfnamefont {A.}~\bibnamefont
  {Lenard}},\ }\href@noop {} {\bibfield  {journal} {\bibinfo  {journal} {J.
  Math. Phys.}\ }\textbf {\bibinfo {volume} {5}},\ \bibinfo {pages} {930}
  (\bibinfo {year} {1964})}\BibitemShut {NoStop}%
\bibitem [{\citenamefont {Zvonarev}(2005)}]{zvonarev_thesis}%
  \BibitemOpen
  \bibfield  {author} {\bibinfo {author} {\bibfnamefont {M.}~\bibnamefont
  {Zvonarev}},\ }\emph {\bibinfo {title} {{Correlations in 1d Boson and fermion
  systems: exact results}}},\ \href@noop {} {Ph.D. thesis} (\bibinfo {year}
  {2005})\BibitemShut {NoStop}%
\bibitem [{\citenamefont {Caux}\ \emph {et~al.}(2007)\citenamefont {Caux},
  \citenamefont {Calabrese},\ and\ \citenamefont
  {Slavnov}}]{2007_Caux_JSTAT_P01008}%
  \BibitemOpen
  \bibfield  {author} {\bibinfo {author} {\bibfnamefont {J.-S.}\ \bibnamefont
  {Caux}}, \bibinfo {author} {\bibfnamefont {P.}~\bibnamefont {Calabrese}}, \
  and\ \bibinfo {author} {\bibfnamefont {N.~A.}\ \bibnamefont {Slavnov}},\
  }\href@noop {} {\bibfield  {journal} {\bibinfo  {journal} {J. Stat. Mech.:
  Th. Exp.}\ ,\ \bibinfo {pages} {P01008}} (\bibinfo {year}
  {2007})}\BibitemShut {NoStop}%
\bibitem [{\citenamefont {Ryu}\ \emph {et~al.}(2007)\citenamefont {Ryu},
  \citenamefont {Andersen}, \citenamefont {Clad\'e}, \citenamefont {Natarajan},
  \citenamefont {Helmerson},\ and\ \citenamefont {Phillips}}]{2007_Ryu_PRL_99}%
  \BibitemOpen
  \bibfield  {author} {\bibinfo {author} {\bibfnamefont {C.}~\bibnamefont
  {Ryu}}, \bibinfo {author} {\bibfnamefont {M.~F.}\ \bibnamefont {Andersen}},
  \bibinfo {author} {\bibfnamefont {P.}~\bibnamefont {Clad\'e}}, \bibinfo
  {author} {\bibfnamefont {V.}~\bibnamefont {Natarajan}}, \bibinfo {author}
  {\bibfnamefont {K.}~\bibnamefont {Helmerson}}, \ and\ \bibinfo {author}
  {\bibfnamefont {W.~D.}\ \bibnamefont {Phillips}},\ }\href {\doibase
  10.1103/PhysRevLett.99.260401} {\bibfield  {journal} {\bibinfo  {journal}
  {Phys. Rev. Lett.}\ }\textbf {\bibinfo {volume} {99}},\ \bibinfo {pages}
  {260401} (\bibinfo {year} {2007})}\BibitemShut {NoStop}%
\bibitem [{\citenamefont {Gring}\ \emph {et~al.}(2012)\citenamefont {Gring},
  \citenamefont {Kuhnert}, \citenamefont {Langen}, \citenamefont {Kitagawa},
  \citenamefont {Rauer}, \citenamefont {Schreitl}, \citenamefont {Mazets},
  \citenamefont {Smith}, \citenamefont {Demler},\ and\ \citenamefont
  {Schmiedmayer}}]{2012_Gring_SCIENCE_14}%
  \BibitemOpen
  \bibfield  {author} {\bibinfo {author} {\bibfnamefont {M.}~\bibnamefont
  {Gring}}, \bibinfo {author} {\bibfnamefont {M.}~\bibnamefont {Kuhnert}},
  \bibinfo {author} {\bibfnamefont {T.}~\bibnamefont {Langen}}, \bibinfo
  {author} {\bibfnamefont {T.}~\bibnamefont {Kitagawa}}, \bibinfo {author}
  {\bibfnamefont {B.}~\bibnamefont {Rauer}}, \bibinfo {author} {\bibfnamefont
  {M.}~\bibnamefont {Schreitl}}, \bibinfo {author} {\bibfnamefont
  {I.}~\bibnamefont {Mazets}}, \bibinfo {author} {\bibfnamefont {D.~A.}\
  \bibnamefont {Smith}}, \bibinfo {author} {\bibfnamefont {E.}~\bibnamefont
  {Demler}}, \ and\ \bibinfo {author} {\bibfnamefont {J.}~\bibnamefont
  {Schmiedmayer}},\ }\href {\doibase 10.1126/science.1224953} {\ \textbf
  {\bibinfo {volume} {337}},\ \bibinfo {pages} {1318} (\bibinfo {year}
  {2012})}\BibitemShut {NoStop}%
\bibitem [{\citenamefont {Penc}\ and\ \citenamefont
  {S\'olyom}(1993)}]{1993_Penc_PRB_47}%
  \BibitemOpen
  \bibfield  {author} {\bibinfo {author} {\bibfnamefont {K.}~\bibnamefont
  {Penc}}\ and\ \bibinfo {author} {\bibfnamefont {J.}~\bibnamefont
  {S\'olyom}},\ }\href {\doibase 10.1103/PhysRevB.47.6273} {\bibfield
  {journal} {\bibinfo  {journal} {Phys. Rev. B}\ }\textbf {\bibinfo {volume}
  {47}},\ \bibinfo {pages} {6273} (\bibinfo {year} {1993})}\BibitemShut
  {NoStop}%
\bibitem [{\citenamefont {Voit}(1995)}]{1995_Voit_RPP_58}%
  \BibitemOpen
  \bibfield  {author} {\bibinfo {author} {\bibfnamefont {J.}~\bibnamefont
  {Voit}},\ }\href@noop {} {\bibfield  {journal} {\bibinfo  {journal} {Rep.
  Prog. Phys.}\ }\textbf {\bibinfo {volume} {58}},\ \bibinfo {pages} {977}
  (\bibinfo {year} {1995})}\BibitemShut {NoStop}%
\bibitem [{\citenamefont {Pletyukhov}\ and\ \citenamefont
  {Gritsev}(2004)}]{2004_Pletyukhov_PRB_70}%
  \BibitemOpen
  \bibfield  {author} {\bibinfo {author} {\bibfnamefont {M.}~\bibnamefont
  {Pletyukhov}}\ and\ \bibinfo {author} {\bibfnamefont {V.}~\bibnamefont
  {Gritsev}},\ }\href {\doibase 10.1103/PhysRevB.70.165316} {\bibfield
  {journal} {\bibinfo  {journal} {Phys. Rev. B}\ }\textbf {\bibinfo {volume}
  {70}},\ \bibinfo {pages} {165316} (\bibinfo {year} {2004})}\BibitemShut
  {NoStop}%
\bibitem [{\citenamefont {de~Vega}\ and\ \citenamefont
  {Woynarovich}(1985)}]{1985_deVega_NPB_251}%
  \BibitemOpen
  \bibfield  {author} {\bibinfo {author} {\bibfnamefont {H.}~\bibnamefont
  {de~Vega}}\ and\ \bibinfo {author} {\bibfnamefont {F.}~\bibnamefont
  {Woynarovich}},\ }\href {\doibase
  http://dx.doi.org/10.1016/0550-3213(85)90271-8} {\bibfield  {journal}
  {\bibinfo  {journal} {Nucl. Phys. B}\ }\textbf {\bibinfo {volume} {251}},\
  \bibinfo {pages} {439 } (\bibinfo {year} {1985})}\BibitemShut {NoStop}%
\bibitem [{\citenamefont {Woynarovich}(1989)}]{1989_Woynarovich_JPA_22}%
  \BibitemOpen
  \bibfield  {author} {\bibinfo {author} {\bibfnamefont {F.}~\bibnamefont
  {Woynarovich}},\ }\href {http://stacks.iop.org/0305-4470/22/i=19/a=017}
  {\bibfield  {journal} {\bibinfo  {journal} {J. Phys. A: Math. Gen.}\ }\textbf
  {\bibinfo {volume} {22}},\ \bibinfo {pages} {4243} (\bibinfo {year}
  {1989})}\BibitemShut {NoStop}%
\bibitem [{\citenamefont {Shashi}\ \emph {et~al.}(2011)\citenamefont {Shashi},
  \citenamefont {Glazman}, \citenamefont {Caux},\ and\ \citenamefont
  {Imambekov}}]{2011_Shashi_PRB_84}%
  \BibitemOpen
  \bibfield  {author} {\bibinfo {author} {\bibfnamefont {A.}~\bibnamefont
  {Shashi}}, \bibinfo {author} {\bibfnamefont {L.~I.}\ \bibnamefont {Glazman}},
  \bibinfo {author} {\bibfnamefont {J.-S.}\ \bibnamefont {Caux}}, \ and\
  \bibinfo {author} {\bibfnamefont {A.}~\bibnamefont {Imambekov}},\ }\href
  {\doibase 10.1103/PhysRevB.84.045408} {\bibfield  {journal} {\bibinfo
  {journal} {Phys. Rev. B}\ }\textbf {\bibinfo {volume} {84}},\ \bibinfo
  {pages} {045408} (\bibinfo {year} {2011})}\BibitemShut {NoStop}%
\bibitem [{\citenamefont {Shashi}\ \emph {et~al.}(2012)\citenamefont {Shashi},
  \citenamefont {Panfil}, \citenamefont {Caux},\ and\ \citenamefont
  {Imambekov}}]{2012_Shashi_PRB_85}%
  \BibitemOpen
  \bibfield  {author} {\bibinfo {author} {\bibfnamefont {A.}~\bibnamefont
  {Shashi}}, \bibinfo {author} {\bibfnamefont {M.}~\bibnamefont {Panfil}},
  \bibinfo {author} {\bibfnamefont {J.-S.}\ \bibnamefont {Caux}}, \ and\
  \bibinfo {author} {\bibfnamefont {A.}~\bibnamefont {Imambekov}},\ }\href
  {\doibase 10.1103/PhysRevB.85.155136} {\bibfield  {journal} {\bibinfo
  {journal} {Phys. Rev. B}\ }\textbf {\bibinfo {volume} {85}},\ \bibinfo
  {pages} {155136} (\bibinfo {year} {2012})}\BibitemShut {NoStop}%
\bibitem [{\citenamefont {Cazalilla}(2004)}]{2004_Cazalilla_JPB_37}%
  \BibitemOpen
  \bibfield  {author} {\bibinfo {author} {\bibfnamefont {M.~A.}\ \bibnamefont
  {Cazalilla}},\ }\href@noop {} {\bibfield  {journal} {\bibinfo  {journal} {J.
  Phys. B: AMOP}\ }\textbf {\bibinfo {volume} {37}},\ \bibinfo {pages} {S1}
  (\bibinfo {year} {2004})}\BibitemShut {NoStop}%
\bibitem [{\citenamefont {Tan}(2008)}]{2008_Tan_AP_323_1}%
  \BibitemOpen
  \bibfield  {author} {\bibinfo {author} {\bibfnamefont {S.}~\bibnamefont
  {Tan}},\ }\href {\doibase http://dx.doi.org/10.1016/j.aop.2008.03.004}
  {\bibfield  {journal} {\bibinfo  {journal} {Ann. Phys.}\ }\textbf {\bibinfo
  {volume} {323}},\ \bibinfo {pages} {2952 } (\bibinfo {year}
  {2008})}\BibitemShut {NoStop}%
\bibitem [{\citenamefont {Imambekov}\ and\ \citenamefont
  {Glazman}(2008)}]{2008_Imambekov_PRL_100}%
  \BibitemOpen
  \bibfield  {author} {\bibinfo {author} {\bibfnamefont {A.}~\bibnamefont
  {Imambekov}}\ and\ \bibinfo {author} {\bibfnamefont {L.~I.}\ \bibnamefont
  {Glazman}},\ }\href {\doibase 10.1103/PhysRevLett.100.206805} {\bibfield
  {journal} {\bibinfo  {journal} {Physical Review Letters}\ }\textbf {\bibinfo
  {volume} {100}},\ \bibinfo {eid} {206805} (\bibinfo {year}
  {2008})}\BibitemShut {NoStop}%
\bibitem [{\citenamefont {Imambekov}\ and\ \citenamefont
  {Glazman}(2009)}]{2009_Imambekov_SCIENCE_323}%
  \BibitemOpen
  \bibfield  {author} {\bibinfo {author} {\bibfnamefont {A.}~\bibnamefont
  {Imambekov}}\ and\ \bibinfo {author} {\bibfnamefont {L.~I.}\ \bibnamefont
  {Glazman}},\ }\href@noop {} {\bibfield  {journal} {\bibinfo  {journal}
  {Science}\ }\textbf {\bibinfo {volume} {323}},\ \bibinfo {pages} {228}
  (\bibinfo {year} {2009})}\BibitemShut {NoStop}%
\end{thebibliography}
\end{document}